\documentclass{pasa}%

\usepackage{graphicx}
\usepackage{caption}
\usepackage{subcaption}
\usepackage{float}
\usepackage{enumitem}
\usepackage{booktabs}
\usepackage{csquotes}
\title{Optimising MWA EoR data processing for improved 21 cm power spectrum measurements - fine-tuning ionospheric corrections}

\author[Chege et al.]{J. Kariuki Chege$^{1,2}$\thanks{Corresponding author: J. K. Chege \newline \href{jameskariuki31@gmail.com}{jameskariuki31@gmail.com} }, C. H. Jordan$^{1,2}$, C. Lynch$^{1,2}$, C. M. Trott$^{1,2}$ J. L. B. Line $^{1,2}$, B. Pindor$^{2,3}$, and S. Yoshiura$^{4,3,2}$
\affil{$^{1}$International Centre for Radio Astronomy Research, Curtin University, Bentley, WA 6102, Australia}
\affil{$^2$ARC Centre of Excellence for All Sky Astrophysics in 3 Dimensions (ASTRO 3D), Bentley, Australia}
\affil{$^3$The University of Melbourne, School of Physics, Parkville, VIC 3010, Australia}
\affil{$^4$Mizusawa VLBI Observatory, National Astronomical Observatory Japan, 2-21-1 Osawa, Mitaka, Tokyo181-8588, Japan}
}

\jid{PASA}
\doi{10.1017/pas.\the\year.xxx}
\jyear{\the\year}

\usepackage{aas_macros}
\usepackage{hyperref}
\usepackage{placeins}
\usepackage[title]{appendix}
\hypersetup{colorlinks,citecolor=blue,linkcolor=blue,urlcolor=blue}

\begin{document}

\begin{frontmatter}
\maketitle

\begin{abstract}
The redshifted cosmological 21 cm signal emitted by neutral hydrogen during the first billion years of the universe is much fainter relative to other galactic and extragalactic radio emissions, posing a great challenge towards detection of the signal. Therefore, precise instrumental calibration is a vital prerequisite for the success of radio interferometers such as the Murchison Widefield Array (MWA), which aim for a 21 cm detection.
Over the previous years, novel calibration techniques targeting the power spectrum paradigm of EoR science have been actively researched and where possible implemented. Some of these improvements, for the MWA, include the accuracy of sky models used in calibration and the treatment of ionospheric effects, both of which introduce unwanted contamination to the EoR window. Despite sophisticated non-traditional calibration algorithms being continuously developed over the years to incorporate these methods, the large datasets needed for EoR measurements requires high computational costs, leading to trade-offs that impede making use of these new tools to maximum benefit. Using recently acquired computation resources for the MWA, we test the full capabilities of the state-of-the-art calibration techniques available for the MWA EoR project, with a focus on both direction dependent and direction independent calibration.
Specifically, we investigate improvements that can be made in the vital calibration stages of sky modelling, ionospheric correction, and compact source foreground subtraction as applied in the hybrid foreground mitigation approach (one that combines both foreground subtraction and avoidance).
Additionally, we investigate a method of ionospheric correction using interpolated ionospheric phase screens and assess its performance in the power spectrum space. Overall, we identify a refined RTS calibration configuration that leads to an at least 2 factor reduction of the EoR window power contamination at the $0.1 \; \text{hMpc}^{-1}$ scale. The improvement marks a step further towards detecting the 21 cm signal using the MWA and the forthcoming SKA low telescope.
\end{abstract}

\begin{keywords}
interferometric – radio telescopes – reionization – techniques: statistical

\end{keywords}
\end{frontmatter}

\section{INTRODUCTION }
\label{sec:intro}
 Low frequency radio interferometers are attempting to observe the cosmic Epoch of Reionisation (EoR) which is one of their main science goals. The EoR is a period after the formation of the first stars and galaxies, before which the early universe was dominated by neutral hydrogen. Radiation from these first sources began `reionising' neutral hydrogen, forming gradually expanding patches of ionised gas around the sources. This was the beginning of the last phase change of the universe from being neutral to its present, almost fully ionised state. Neutral hydrogen emits a spectral line signal at 1420 MHz (21 cm) and it's this signal, from the EoR, that radio interferometers such as the Murchison Widefield Array (MWA, \citealt{Bowman2013, Tingay2013}), Long Frequency Array (LOFAR, \citealt{VanHaarlem2013}) and the Hydrogen Epoch of Reionisation Array (HERA, \citealt{Deboer2017}) are targeting. The 21 cm signal promises to be the best probe of this era and its observation would provide comprehensive answers to many questions pertaining to properties of the first galaxies, improving our knowledge on the  evolution of the universe as a whole (see reviews in, \citealt{Fan2006, Furlanetto2006, Pritchard2012, Zaroubi2013}).

Current interferometers lack the sensitivity required to directly detect the cosmological signal, estimated to be a few tens of milliKelvin in brightness temperature. A more feasible pursuit has been targeting a statistical detection of the signal, with most efforts going towards measuring its power spectrum (PS).  Further, radiation from emitters (foregrounds) that are 2 to 3 orders of magnitude brighter, make detection of the signal an uphill task. Deriving ways of mitigating these foregrounds has remained a key area in observational EoR research.
 
Most current foreground mitigation methods hinge on the contrast between the spectral signature of the foregrounds and the 21 cm signal. In contrast to the foregrounds, which are mostly spectrally smooth, the 21 cm signal varies rapidly over frequency.
Therefore, in Fourier ($k$) space, the 21 cm signal and the foregrounds occupy separate distinguishable $k$ modes. The 21 cm signal dominates the foreground-free modes and in principle, with enough sensitivity, is detectable.
The MWA foregrounds mitigation strategy is a hybrid between the foreground avoidance strategy and foreground subtraction, the two main methods employed in the field. In foregrounds avoidance, foreground-dominated modes (the wedge feature in 2D PS, \citealt{ Datta2010, Parsons2012, Vedantham2012, Morales2012}) are discarded. On the other hand, foreground subtraction attempts to directly remove a model of the unwanted sky from the data by use of either parametric or non-parametric methods, leaving the cosmological signal behind. So far, specifically for the MWA, foreground subtraction alone has not been successful in helping recover the foreground dominated modes. However, \citealt{Kerrigan2018} showed that such a hybrid method can result in improved results for the modes in the EoR window. In this hybrid mitigation method, the main benefit of foregrounds subtraction is realised from the fact that by subtracting power from the wedge, we are by extension reducing the power that can leak to the window and therefore keeping the window as contamination free as possible for a 21 cm detection \citep{Liu2020}.
Foregrounds necessitate precise instrumental calibration, with the success of any foreground mitigation strategy relying on a calibration that achieves a challenging dynamic range of $\sim10^5$.

Spectral calibration errors corrupt power in $k$ space. Over the last few decades, instrumental calibration challenges have led to a push for non-traditional calibration techniques specific for EoR science with interferometers. Using simulations, \citealt{Barry2016, Byrne2019} investigated the effect of incomplete sky models in EoR calibration and found that errors due to faint unmodelled sources in sky models results in erroneous calibration solutions that contaminate the EoR window and could inhibit a 21 cm detection. Errors in positions of the sources themselves will also introduce unwanted calibration errors and as a result, custom source catalogues for enhanced accuracy during MWA sky modelling are used \citep{line2017}. There are also continuous efforts to include a diffuse emission component in EoR sky models \citep{Bryne2021}, improved radio frequency interference (RFI) mitigation \citep{Wilensky2019}, as well as improving the accuracy of the instrumental beam response \citep{Sokolowski2017}. 

We focus on the ionosphere as a cause of calibration errors firstly because for the MWA, ionospheric refraction, which results in positional offsets of compact sources, is a dominant effect. The ionosphere manifests as a direction dependent effect based on the MWA array properties; see \cite{lonsdale2005}. MWA PS measurements usually exclude data observed in durations with high ionospheric activity \cite{Trott2018, Trott2020, Rahimi2021, Yoshiura2021}.
Secondly, the Real-Time System (RTS, \citealt{Mitchell2008}), the calibration algorithm used in this work, is designed to perform a thorough direction dependent calibration, focusing almost entirely on the ionosphere.
Since the primary ionospheric effects are expected to manifest in the foreground dominated $k$ modes, using simulations, this paper shows how ionospheric refraction leads to contamination in $k$ modes that otherwise should be clean and also test whether increased sky model sources combined with aggressive foregrounds subtraction helps in reducing the contamination.
The primary effects due to the ionosphere are expected to manifest in the foreground dominated $k$ modes. Using simulations, this paper shows how ionospheric refraction leads to ionospheric contamination leaking into $k$ modes that otherwise should be clean. In addition, we test whether increased sky model sources applied together with aggressive foregrounds subtraction helps in reducing the contamination.

\citealt{Yoshiura2021} has shown the effect of ionospheric errors at MWA `ultra-low' (below 100 MHz) frequencies, which result in poor direction-independent calibration solutions. This shows that the ionosphere also does detrimentally affect calibration, analogously to a sky model that has source positional errors. To further reduce these ionospheric related calibration errors, they showed that an additional direction-indpendent (DI) calibration step, which now uses an updated sky model with ionospheric source positional offsets accounted for, improved calibration solutions before repeating the direction-dependent (DD) calibration step. We test whether this method of modifying sky models based on measured ionospheric offsets reduces contamination in the higher MWA frequency bands ($\sim 140$-$170$ MHz, MWA). In addition, we investigate whether more accurate ionospheric modelling techniques are helpful for MWA calibration. We use these tests to determine the best method for calibrating observed MWA data.

Despite ongoing development of EoR-specific calibration algorithms that implement the aforementioned new non-traditional calibration routines that would potentially improve the MWA PS measurements, a full implementation and testing of these methods has not been feasible over the previous years. The task involves high computational costs, exacerbated by the need for EoR studies to average over potentially thousands of hours of data to reach the required sensitivity.

In 2020, the MWA acquired \textsc{garrawarla}, a supercomputer dedicated to MWA data processing hosted by the Pawsey Supercomputing Centre.
With this recent significant upgrade in our processing capabilities, we are able to perform these tests and enhance our calibration knowledge, a big motivation for this work. This work will feed into other ongoing software development efforts; more efficient and faster tools targeting the MWA and the forthcoming Square Kilometre Array (SKA).

The main objective of this work is to identify the most accurate calibration routine for MWA EoR that fully utilises the capabilities of the currently available calibration tools, specifically the RTS, with a focus on ionospheric correction, foregrounds subtraction and sky model completeness.

This paper is organised as follows: In Section \ref{method}, we describe the methods applied and the data used. This involves our simulated models for the ionosphere and the real data used, as well as the calibration routine and PS estimation applied in this work. We then introduce the MWA real observations dataset and the validation procedure adopted. Section \ref{results} presents the results, and Section \ref{sfinterp} investigates more methods of ionospheric modelling. A discussion and conclusion of the work is given in Section \ref{sec:discuss}.

\section{Analysis methods}
\label{method}
\subsection{Visibilities and calibration formalism}
Radio interferometers measure data referred to as visibilities, computed using the visibility measurement equation \citep{Thompson2017}.

\begin{equation}
\boldsymbol{v}^{meas}=\int\int{\frac{A(l,m)I(l,m)}{\sqrt{1-l^{2}-m^{2}}}e^{-2\pi{i}\phi}} dl \, dm,
\label{ms_eqn}
\end{equation}
where
$$\phi = ul+vm+w(\sqrt{1-l^{2}-m^{2}}-1).$$ 
$(u, v, w)$, also implicit in $\boldsymbol{v}^{meas}$, is a Cartesian coordinate system used to describe the baselines, with $w$ pointing to the direction of the phase centre and $(l, m)$ are the directional cosines. $I(l,m)$ and $A(l,m)$ represents the sky brightness distribution, and the instrumental response as a function of the effective collecting area and the direction of the incoming signal respectively. Assuming coplanar antenna locations ($w=0$) or a small field of view ($\sqrt{1-l^{2}-m^{2}}\approx1$), the Van Cittert Zernike theorem relates the measurement equation to the 2D Fourier transform of the sky. However, multiple systematics corrupt the signal in its propagation path, resulting into $\boldsymbol{v}^{meas}$ discrepant from the true sky visibilities $\boldsymbol{v}^{true}$. Calibration is the process that aims to correct for this discrepancy using a myriad of methods that revolve around minimising the difference between an estimate model of $\boldsymbol{v}^{true}$ and $\boldsymbol{v}^{meas}$ to obtain correction factors, typically referred to as calibration gain solutions or simply gains. See \citealt{Smirnov2011a} and their associated works for a full description on calibration. The calibration step is crucial as it dictates not only the credibility of the subsequent PS estimation step but also how close the output PS upper limits are to the 21 cm signal level. Since the RTS calibration software is a key component of the analysis presented in this work, a distinction between the direction-dependent (DD) and -independent (DI) calibration steps, as implemented by the RTS, is of importance to this work.

\subsection{RTS calibration}
\label{calibration}
The RTS \citep{Mitchell2008} is a GPU-based calibration software package designed for the MWA and is used extensively for MWA EoR observations. It can optionally average visibilities, flag RFI, perform a sky-model based DI calibration, as well as a DD calibration stage that is targeted towards performing an elaborate ionospheric correction routine. Additionally, The RTS can also perform source subtraction to provide residual visibilities from which the PS can be estimated. For both DI and DD calibration, we exclude baselines shorter than 20 wavelengths and apply a taper to remaining baselines less than 40 wavelengths. These shortest baselines are most sensitive to the large scale galactic emission, which is not included in our calibration sky models. Inclusion of these short baselines without a diffuse calibration sky model have been known to bias calibration outputs, see \citep[e.g.][]{Patil2016}

\subsubsection{RTS DI calibration}
\label{patch}
DI calibration aims to solve for the frequency and amplitude gains that are instrumental in nature and therefore unaffected by the direction of observation. The MWA EoR team has a custom-made sky model, matched up using \textsc{puma} \citep{line2017}, from several low frequency surveys which is used to perform an in-field calibration using the RTS. For each observation to be calibrated, the brightest sources after primary beam attenuation within a given sky radius centred on the observed sky field are chosen and modelled together to be used as a single super-calibrator-source. The model visibilities of these sources are used to obtain the gains on the observed visibilities. 

\subsubsection{RTS DD calibration}
\label{peel}
DD calibration corrects for effects that are non-uniform across the sky (i.e. directional), such as the ionosphere. The RTS is designed to apply an iterative per-source DD calibration, with an end result of calibrated foreground subtracted visibilities ready for PS estimation. This is done through an application of the peeling algorithm \citep{Noordam2004}, to a few dominant sources ($\sim5$) followed by an ionospheric calibration, then a direct subtraction of a specified number of sources ($\sim1000$) from the visibilities. However, \citealt{Yoshiura2021} found that applying the full peeling process to the 5 brightest or fewer sources induced calibration errors that resulted in higher power spectrum contamination than doing no peeling at all. Therefore, similar to their work, we omit the peeling procedure entirely and only do the ionospheric calibration process followed by source subtraction. The ionospheric calibration step is further summarised below.

The total beam-attenuated visibilities accumulated from a pool of the brightest calibrator sources to be corrected for the ionosphere are first subtracted from the data in a step referred to as prepeeling. Prepeeling allows for calculation of ionospheric gains in the direction of each source individually, without sidelobes confusion from other sky directions. In descending order of brightness, the sources are one by one re-added back into the residual visibilities and the catalogued position of the source is rotated to be at the phase centre of the observation. This rotation implies that for a source in its correct sky position in the data, the $(l,m)$ values from equation \ref{ms_eqn} should be zero. Any offsets for the source per frequency channel are then fitted for the $\lambda^2$ spectral signature of the ionosphere, and a model of ionospheric refraction on the source is obtained; see Equation 5 in \citealt{Mitchell2008}.The gains obtained for the source direction are primarily used in subtracting it from the data, and the process is repeatedly carried out for all the remaining calibrators.

\subsection{PS estimation}
\label{ps_estimation}
The power spectrum, $P(k)$, is the Fourier transform of the 2-point correlation function (Equation \ref{spintemp}, e.g., \citealt{Zaldarriaga2004}) and it probes the fluctuations of the 21 cm brightness temperature along the line of sight (frequency modes) as well as spatially (angular modes).

\begin{equation}
\langle T_b(\vec{k})\tilde{T}_b(\vec{k}^{\prime})\rangle = (2\pi)^3\delta(\vec{k}-\vec{k}^{\prime})P(k),
\label{spintemp}
\end{equation}

where $\langle\rangle$ indicates the ensemble average and $\delta$ is the Dirac delta function. Interferometers are natural PS measuring instruments, and the PS can be computed directly from the visibilities. We use estimates of the PS to compare the performance of the different experiments carried out. Specifically, we apply the gridded visibilities-based PS estimation approach, which can be summarised by the following steps:

\begin{enumerate}[label=(\alph*)]
\item The visibilities are first gridded onto the ($u$, $v$) plane per frequency channel. A spatial Blackman-Harris gridding kernel was used in this work (use of an instrumental beam kernel has been preferred in the past, but \citealt{Barry2022} recently  showed that use of non-instrumental kernels in this step does not introduce significant errors.)
\item To smooth the step function resulting from the discrete frequency bands, a taper, usually Blackman-Harris for the MWA, is often applied.
\item A Fourier Transform is then taken in the Frequency direction and the result squared to obtain the PS.
\end{enumerate}

The PS obtained from the steps above is 3 dimensional, and with ($u$, $v$, $\eta$) dimensions where  where $\eta$ is the Fourier dual to frequency.
It is common practice to compute its weighted averages in either cylindrical or spherical shells to obtain the 2D or 1D PS versions, respectively.
The 2D PS has the distinct EoR window and foregrounds wedge morphology, with axis units of $k_{\perp}=\sqrt{(k_u^2+k_v^2)}$, $k_{\parallel} \propto \eta$ obtained from the two ($u,v$) baseline directions and the line-of-sight (frequency) components respectively.
The occurrence of these features is due to the mode mixing phenomenon \citep{Datta2010, Liu2011, Parsons2012, Vedantham2012, Morales2012}. Mode mixing results from spectrally smooth foreground components that lie at small $k_{\parallel}$ modes interacting with the instrument chromaticity. As a result, mode mixing leaks power to higher $k_{\parallel}$ as a function of $k_{\perp}$ resulting in the foregrounds wedge feature.
The EoR window remains as a region with lower foregrounds power and the most promising for detection of the 21 cm signal. The 2D PS is a crucial diagnostic in examining how different aspects of the 21 cm experiment lead to varying power in both the wedge and the window modes.
The 1D PS on the other hand is mostly used to draw the power limits and is usually presented as a dimensionless quantity by integrating the total power on a given spatial scale over the volume.
A full description together with the conversions from $k$ to cosmological dimensions  can be found in \citealt{Morales2004, McQuinn2006}.

In this work, we use both the 1D and 2D PS computed using the Cosmological HI PS Estimator (\textsc{chips}, \citep{Trott2016}) to describe our results. CHIPS is optimised for estimation of MWA data power spectra and has been used in previous literature (e.g. \citealt{Trott2020, Rahimi2021, Yoshiura2021}).

\begin{table*}[th!]
\begin{center}
\caption{MWA observational and ionospheric simulation parameters.}
\begin{tabular}{ l l l }
\hline
\hline
Phase Centre (RA, Dec; J2000) & $0.0^\text{h}$, $-27^{\circ}$ &  \\ 
Minimum Frequency & 138.8 & MHz \\  
Maximum Frequency & 169.5 & MHz \\
Full Bandwidth  & 30.7 & MHz \\
Duration          & 112   & s   \\
\hline
Sub bands & 24 & \\
Sub-band width & 1.28 & MHz \\
Fine channels per sub-band & 32 & \\
Total fine channels & 768 & \\
Fine channel width & 40 & kHz \\
Total timestamps & 14 &  \\
Duration per timestamp & 8 & s \\
\hline
Spatial ionospheric structure & $K$, $S$ & \\
Turbulence level hyperparameter & 5 - 300 &  \\
Offsets magnitude & 0.01 - 0.5 & arcmin
\end{tabular}
\label{tab:sim_params}
\end{center}
\end{table*}

\section{Simulations}
\subsection{Ionospheric modelling}

\begin{figure*}[!th]
    \centering
    \includegraphics[width=\textwidth]{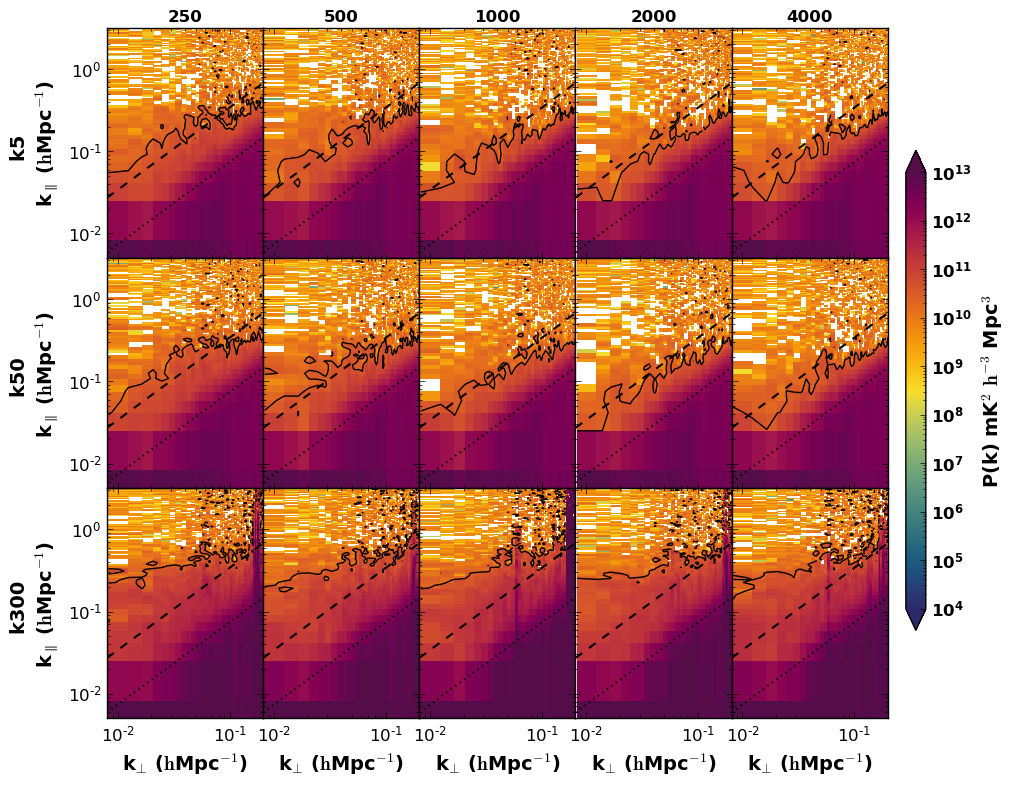}
    \caption{2D power spectra for 5000 sources simulations with Kolmogorov-like ionospheric models. The rows indicate increasing ionospheric turbulence levels from top to bottom.
    For each panel, the same amount of sources is used for both DI and ionospheric calibration; indicated at the top of each column and increasing from left to right. A constant 1000 sources were subtracted from the calibrated visibilities in all panels.} The black solid contour shows the $10^{10.5}$ mK$^2$h$^{-3}$Mpc$^3$ power level. The EoR window is noise-dominated even for the best ionospheric conditions and calibration. For this reason, we draw conclusions of ionospheric impact using lower noise simulations shown in Figure \ref{fig:k5000}. The dashed and dotted lines represent the `horizon' and the `primary field of view' foreground power limits, based on source positions in the sky and the MWA primary beam, respectively.
    \label{fig:k5000noise}
\end{figure*}

\begin{figure*}[!ht]
    \centering
    \includegraphics[width=.87\textwidth]{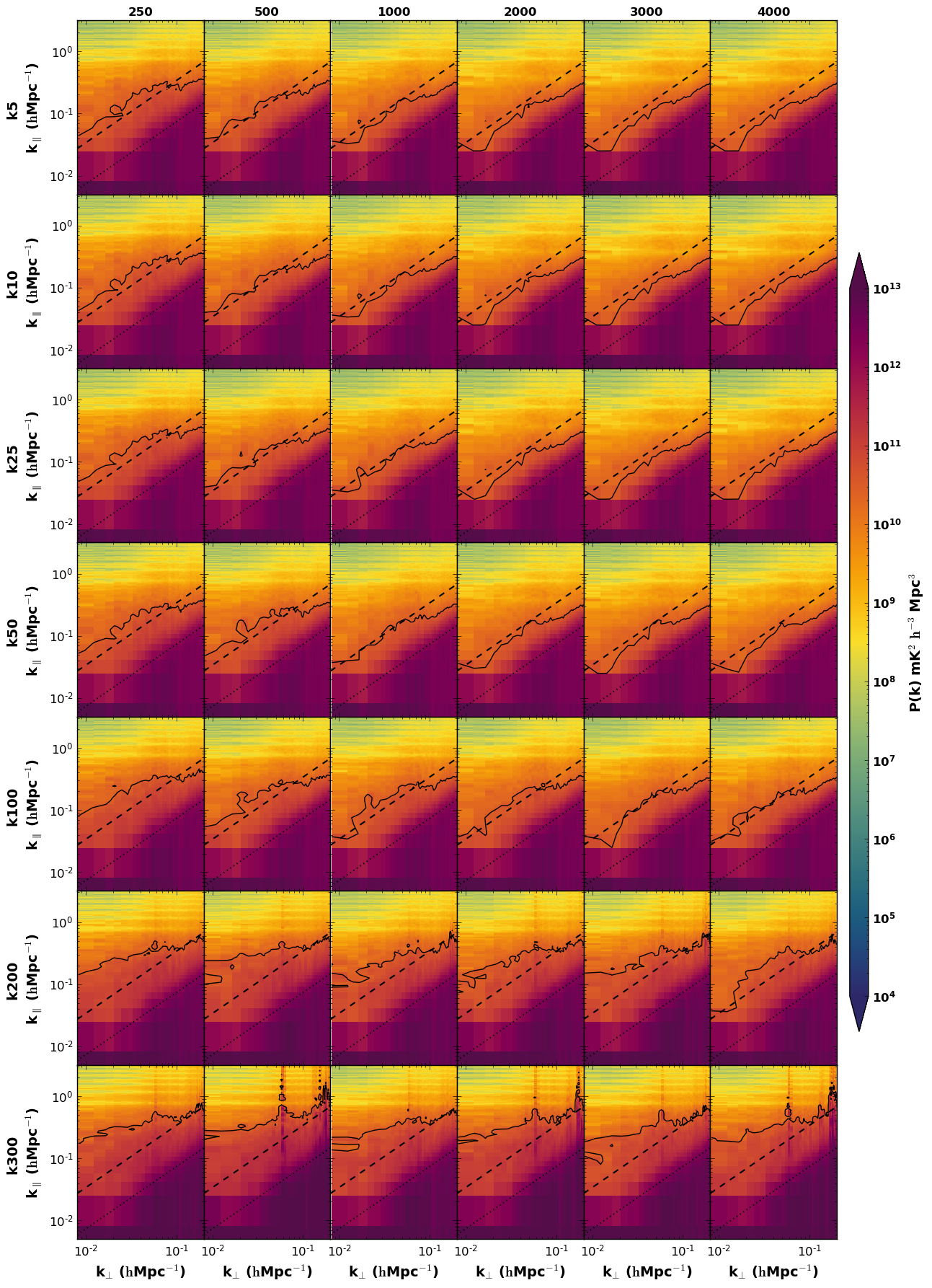}
    \caption{2D power spectra for 5000 sources simulations with Kolmogorov-like ionospheric models and scaled down thermal noise. The rows indicate increasing ionospheric turbulence levels from top to bottom. For each panel, the same amount of sources is used for both DI and ionospheric calibration; indicated at the top of each column and increasing from left to right. A constant 1000 sources were subtracted from the calibrated visibilities in all panels. The black solid contour shows the $10^{10.5}$ mK$^2$h$^{-3}$Mpc$^3$ power level. Increasing the number of sky model sources in all calibration steps reduces power in all power spectrum modes. The most improvement is obtained when the ionosphere is most inactive.}
    \label{fig:k5000}
\end{figure*}

When including positional offsets induced by the ionosphere, an ionospheric differential phase offset, $\phi'$, term evaluated between two lines of sight through the ionosphere from each antenna pair in the array is added to the $\phi$ term in equation \ref{ms_eqn}.
To model this effect in mock MWA visibilities, we use the \textsc{sivio} package \citep{chege2021}, which makes thin phase screen models of ionospheric refraction as well as MWA parameters to compute the visibilities in Equation \ref{ms_eqn}. We modelled phase screens with Kolmogorov ($K$) and sine-like ($S$) spatial structures. The $K$ screen can be described by an isotropic 2-dimensional (2D) Gaussian random field with a power law power spectrum of the form $\sim|\mathbf{k}|^{11/3}$ where $\mathbf{k}$ is the 2D Fourier wavenumber \citep[e.g.,][]{Vedantham2016}. The $S$ screen is described by a smooth-varying 2D sine function of the form $\gamma\times\text{sin}\left(\sqrt{\alpha x^2+ \beta y^2}\right)$ with $(x,y)$ being the two axis of the screen, $\gamma$ modifying the number of `ridges' and ($\alpha, \beta$) modulating their shape. 

The ionospheric turbulence level is modulated by a \textsc{sivio} magnitude hyperparameter, which can be interpreted as a scaling factor on the simulated ionospheric differential total electron density. In this work, 7 turbulence levels of $5, 10, 25, 50, 100, 200$ and $300$ were simulated. The values signify ionospheric activity levels, ranging from very calm (level $5$; $	\lesssim 0.01$ arcmin source position offsets) to an extremely active (level $300$; $\sim 0.5$ arcmin source position offsets). These turbulence levels are comparable to the ones observed in real MWA data by \citealt{jordan2017}. The simulated visibilities assume a static ionosphere in $2$ minutes durations, subdivided into $14$ separate time stamps, 8s each. The simulations are of $30.72$ MHz total bandwidth, composed of 768 fine channels of $40$ kHz width each between $\sim 140$ - $170$ MHz, a replica of real MWA low band data. The simulated data provides ionospheric effects that are dominant over any other systematics therefore testable.

\subsubsection{Foregrounds and noise}
We centre the simulated data at RA$=0^{\circ}$ and Dec=$-27^{\circ}$ (EoR0), which is one of the main fields used for MWA EoR observations, as it is located at a cold sky patch away from the radio loud galactic plane and other bright extended sources. Our sky model is composed of the 5000 brightest sources obtained from the GLEAM catalogue within a 20 deg sky radius. This region is fully encompassed by the main lobe of the MWA beam response at the low band. We do not include any wide field\footnote{The primary lobe of low frequency arrays like the MWA is considered to be \textit{widefield}. Here we refer to not including sources located on the sidelobes all the way to the horizon as real data MWA calibration models would.} simulations or sources with extended morphologies whose ionospheric offsets are challenging to quantify accurately.
All the sources are modelled as point sources shifted from their catalogue locations by the respective ionospheric activity along the different paths from the array elements to the source. Modelling the sources as compact point sources allows for more accurate measurements of their positional shifts.

Based on the use case, as described in the following sections, we also add thermal radiometric measurement noise estimated as:

\begin{equation}
    \label{thermal}
    \sigma_N = 10^{26}\frac{2k_\text{B} T_{\text{sys}}}{A_{\text{eff}}}\frac{1}{\sqrt{2\Delta\nu\Delta t}} \;\;\; \text{Jy}, 
\end{equation}

where $T_{\text{sys}}$ is the system temperature, $A_{\text{eff}}$ is the effective collecting area of the antenna, $\Delta\nu$ is the frequency resolution, and $\Delta t$ is the integration time of the visibilities. Typical values for the MWA are $T_{\text{sys}} = 240$ K, $A_{\text{eff}}=21$ m$^2$, $\Delta\nu = 40$ kHz and $\Delta t = 8$ s.

As ionospheric effects are best studied using compact sources, we do not add any large-scale sky model components such as the galactic diffuse emission to the visibilities. Since the EoR signal is relatively weak when compared to the above foregrounds, we leave it out of the visibility simulations.

\subsection{Key questions and analysis procedure}
Having laid out the relevant background, we now summarise the specific questions this work aims to investigate:
\begin{enumerate}[label=(\alph*)]
\item To what extent do ionospheric effects manifest in the EoR window?
\item Do different ionospheric structures result in different levels of contamination?
\item How many sky model sources are optimum for the DI, and DD RTS calibration as well as for source subtraction?
\item Does performing a sky model correction based on measured ionospheric offsets result in lower PS contamination for the MWA low band?
\item What is the ultimate best calibration routine for the RTS at the low band with respect to the ionosphere?
\end{enumerate}

A summary of the experiments run on the simulated data is described below:

\begin{enumerate}
    \item We compute the Kolmogorov and Sine-like spatial ionosphere screens. Visibilities for each ionosphere kind are computed for a 2 minutes duration, with turbulence levels ranging from 5 to 300. The exact simulation parameters are summarised in Table \ref{tab:sim_params}
    \item For each observation, we compute two sets of visibilities; one with thermal noise added and the other with a scaled down thermal noise level.
    \item For both sets of simulated data, we vary the number of DI and ionospheric calibrators ranging from 250 to 4000 sources, while keeping the amount of subtracted foregrounds constant (1000 sources,  Figures \ref{fig:k5000noise} and \ref{fig:k5000}).
    
    \item We test whether the RTS ionospheric correction improves calibration (Figure \ref{fig:k100_ionocal_test}).
    
    \item In addition to the number of DI and ionospheric calibrators, we also test the effect of varying the amount of foregrounds subtracted (Figure \ref{fig:ptvspl_k}).
    
    \item A PS is then computed for each individual observation.

\end{enumerate}

\section{Simulation results}
\label{results}

\subsection{Simulations with thermal noise}
\label{pswithnoise}
Figure \ref{fig:k5000noise} shows a suite of 2-minute MWA simulations, calibrated and used to obtain the 2D PS as fully described in Section \ref{method}. The number of DI and ionospheric calibration sources used during calibration are equal for each run but increasing across the columns from 250 to 4000, while the ionospheric turbulence increases for each row from top to bottom. For all panels, a constant 1000 sources was subtracted after DD calibration. The third column is synonymous with the parameters (1000 sources for both DI and DD) used for calibration in the recent best limits results by \citep{Trott2020}.
The first two rows show ionospheric offsets within 0.15 arcmin and therefore, in real data, they would be used for EoR power estimation as done in \citealt{Yoshiura2021}.
To date, we have not established a maximum cutoff of acceptable ionospheric contamination when aggregating data for PS estimation. We expect that, in general, less ionospheric activity is better. However, we wish to better understand how poor ionospheric conditions are allowed to be before we discard data, to find the optimum compromise.

For clarity, the black solid contour shows the $10^{10.5}$ mK$^2$h$^{-3}$Mpc$^3$ power level. As the number of calibrator sources is increased, the power level recedes towards lower $k_{\parallel}$ modes. This improvement is seen even under very calm ionospheric conditions (first row) with the rate of EoR window improvement declining inversely with ionospheric activity.  However, for a single 2-minute observation, Figure \ref{fig:k5000noise} shows that the EoR window is thermal noise dominated even with the best ionospheric conditions and calibration. Therefore, for the ionospheric effects to manifest without many hours of averaged data, we decide to reduce the thermal noise in the simulations.

\begin{figure}[!h]
    \centering
    \includegraphics[width=.5\textwidth]{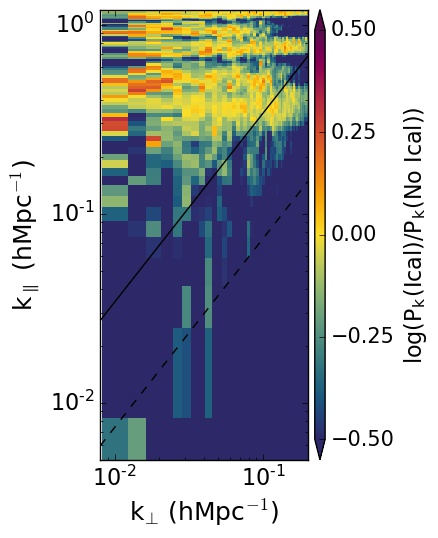}
    \caption{The log of the ratio between the 2D PS with and without RTS ionospheric correction for a $k100$ simulation and scaled down noise. The dominant blue colour indicates that ionospheric correction does indeed reduce contamination due to ionospheric activity.}
    \label{fig:k100_ionocal_test}
\end{figure}

\begin{figure*}[!ht]
    \centering
    \includegraphics[width=\textwidth]{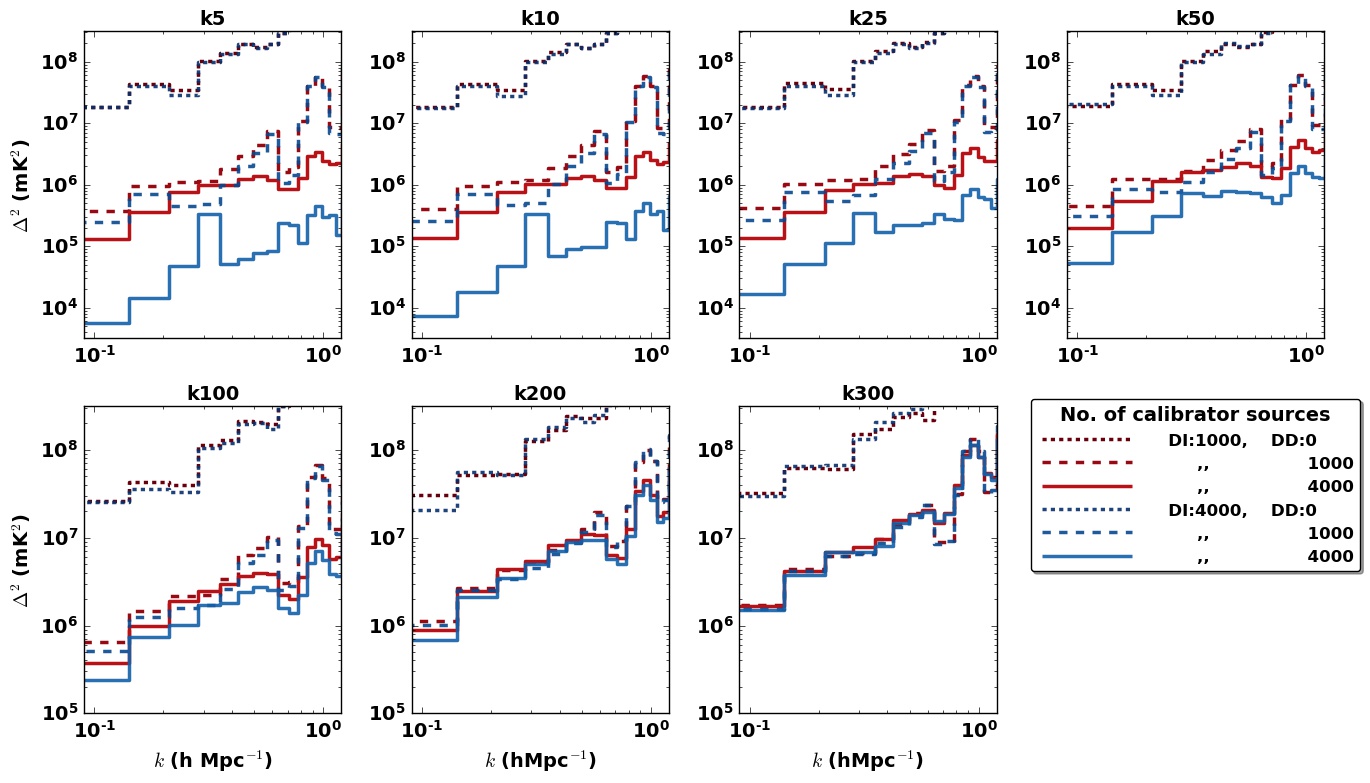}
    \caption{EoR window 1D PS comparison of varying the number of sources in DI and DD with $k$-screen ionospherically contaminated visibilities. The simulation is composed of 5000 point sources with scaled down noise levels. Here, the DD value represents the number of sources that were both ionospherically corrected and subtracted. This figure shows how a combination of inactive ionospheric conditions, more complete DI calibration sky models and subtraction of more compact foregrounds results in reduced contamination in the EoR window.}
    \label{fig:ptvspl_k}
\end{figure*}

\subsection{Low noise simulations}
\label{iono_correction}
We first demonstrate the benefits of using more sources in calibration by using simulations with a minimal thermal noise level.
Figures \ref{fig:k5000} is similar to Figure \ref{fig:k5000noise} but with more individual runs included and with the noise in each observation scaled down by a factor of $10^4$. The effect of both the ionospheric activity as a function of the calibrator sources is more evident. More DI and DD sources result in lower power levels for all ionospheric conditions. As we approach the top right panel, where the ionosphere is most inactive, and with maximum amount of calibration sources, less and less power bleeds to the EoR window. As expected, the bottom left represents the extreme opposite. For the most extreme ionosphere, the foreground subtraction results in structured residuals around sources due to a highly distorted point-spread function (see \citealt{Koopmans2010}). The structured artefacts, especially around the brightest sources, appear as the vertical high-power stripes visible in the two bottom rows of Figure \ref{fig:k5000}.

\subsubsection{RTS ionospheric correction}
To validate the effectiveness of ionospheric correction with the RTS, we run a simulation with scaled down noise and with a $K100$ ionosphere. This data was then processed twice in an identical procedure, except that the first processing run ($r_1$) applied RTS ionospheric correction while the second processing run ($r_2$) did not. In Figure \ref{fig:k100_ionocal_test}, we present the log of the ratio of the 2D power spectra obtained from the two runs ($log(r_1/r_2)$). The dominant negative ratio (bluer hue) indicates less power in the first run, evidence that the RTS ionospheric correction does help reduce ionospheric contamination. Furthermore, the less contamination is not only observed in the wedge alone, but extends to the EoR window as well. However, the ionospheric correction seems to marginally introduce additional contamination in some regions of the window. This could be caused by any spectral errors introduced by the ionospheric correction procedure coupling with the known MWA harmonic systematics. The harmonic systematics result from intermittent course band flagging which is done at regular frequency intervals and manifest as ridges of enhanced power on the PS, especially towards high $k_\parallel$. Additional errors at high $k_\parallel$ could be introduced by the krigging process applied by CHIPS in an attempt to get rid of the coarse band harmonics \citep{Trott2020}.

\subsubsection{Comparing number of DI vs DD sources}
\label{patch_vs_peel}
To better understand the effect of using variable combinations of sky model calibrator source quantities in the DI and DD steps, we run the scaled down noise simulations using 1000 and 4000 DI calibrator sources and for each DI number, we perform DD (ionospheric correction and subtraction) calibration runs with 1000 and 4000 sources. For comparison, Figure \ref{fig:ptvspl_k}
shows the 1D PS computed for each combination of the DI and DD runs using a $3.5k_{\parallel}>k_{\perp}$ cut in order to consider only the window modes. In each panel, and for both the 1000 and 4000 DI runs, we also plot the power spectrum with no DD calibration applied. The red dashed and blue solid lines show the runs with 1000 and 4000 sources, respectively, in both the DI and DD calibration. This is a direct comparison between how all previous MWA data has been processed and a quadruple in the number of calibrator sources. There is at least an order of magnitude difference in power between the two cases for all ionospheric cases except the 3 most extremes ones starting from $k100$ ionosphere. Similarly, omitting the DD calibration and subtraction step altogether (dotted lines) results in up to 3 orders of magnitude difference from the best run (solid blue). The dotted lines further show that without DD calibration, a better DI model does not provide any significant improvements in the power spectrum. Conversely, with the DI calibrators kept constant, more DD calibrators results in significantly lower power.

The red solid line shows the run with 1000 DI sources and 4000 DD sources, while the blue dashed line represents the reverse (4000 DI sources and 1000 DD sources). In almost all ionospheric cases, these two RTS configurations show similar power levels, but the former shows slightly less power in some modes. This indicates how increased calibrator sources are important in not just one step of calibration, but in both the DI and the DD for maximum improvements to be realised.

Higher ionospheric levels result in higher 1D power. For the most extreme ionospheres (last two plots, k200 and k300, note the different y-axis range. Similarly observed for s200 and s300.), the calibration performance has become extremely poor and all the RTS DD configurations make no difference. This is also accompanied by an increased power spectrum slope, resulting from failure by the RTS to accurately subtract extremely offset sources.

The finding of the simulation runs can be summarised as follows:
\begin{enumerate}
    \item Skipping DD calibration and foregrounds subtraction or using fewer DD sources results in more overall power in all modes.
    \item Higher ionospheric turbulence levels result in higher overall power levels.
    \item Without DD calibration, a better DI calibration model results in only marginal improvement.
    \item As the ionosphere gets worse, its induced errors result in more power dominating all the DI and DD errors.
\end{enumerate}

Ideally, source subtraction should only reduce power in the wedge. However, a key observation from the simulations is that the RTS calibration process is not perfect; regardless of configuration, power leaks into the EoR window from the wedge. By subtracting power from the wedge, we are by extension reducing the power that can leak to the window and helping to keep the window as contamination free as possible for 21 cm detection.

For the simulated data, no significant difference in power is observed between the $K$ and the $S$ screens (see Figure \ref{fig:ptvspl_s} for the S-screen analogue of Figure \ref{fig:ptvspl_k}). The reason for this observation has not yet been established and will warrant more investigation in future. Similarly, in the other simulation experiments carried out in this work, no significant result differences were observed from using either a $K$ or $S$-screen.

\section{MWA Real data}
\label{real_data}

\begin{figure}[!th]
    \centering
    \includegraphics[width=.5\textwidth]{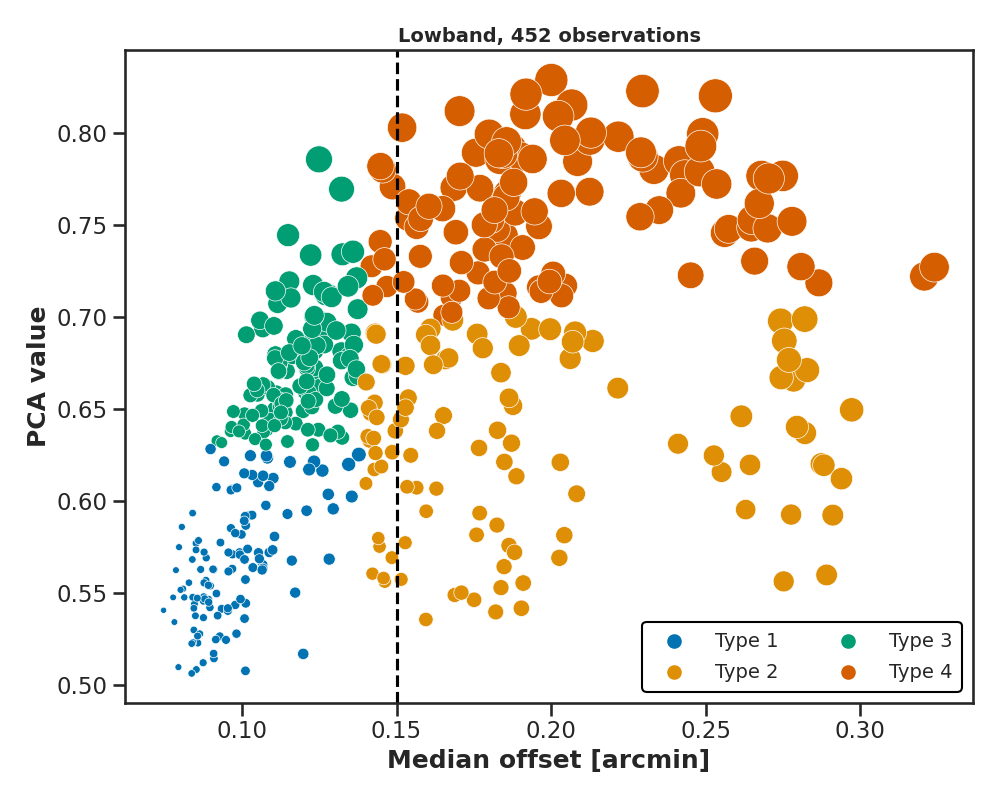}
    \caption{Summary of 452 real data observations used in the analysis. The dashed line shows a typical data cut, where only observations with lower median offsets are analysed further. A distribution of different ionospheric conditions observed within a 1 hour LST interval (LST 0hrs) was targeted. The marker size shows increasing ionospheric QA metric; a linear combination of the two axes values. The PCA value signifies spatial anisotropy in the ionosphere. The observations were carried out in 2014 and 2015.}
    \label{fig:obsids}
\end{figure}

\begin{table*}
\begin{center}
\begin{tabular}{|l|l|l|l|}
\hline
  & 1  & 2  & 3                                                                                                                                      \\ \hline \hline
A & 1000 DI                                                                                             & 4000 DI                                                                                             & N/A                                                                                                    \\ \hline
B & \begin{tabular}[c]{@{}l@{}}1000 DI\\ 0 ionocal\\ 1000 subtracted (DD)\end{tabular}                             & \begin{tabular}[c]{@{}l@{}}4000 DI\\ 0 ionocal\\ 4000 subtracted (DD)\end{tabular}                             & \begin{tabular}[c]{@{}l@{}}4000 DI\\ 0 ionocal\\ Bright sources subtracted (DD)\end{tabular}                                             \\ \hline
C & \begin{tabular}[c]{@{}l@{}}1000 DI\\ 1000 ionocal\\ 1000 subtracted (DD)\end{tabular}                          & \begin{tabular}[c]{@{}l@{}}4000 DI\\ 4000 ionocal\\ 4000 subtracted (DD)\end{tabular}                             & \begin{tabular}[c]{@{}l@{}}4000 DI\\ Bright sources ionocal\\ 4000 subtracted (DD)\end{tabular}                                          \\ \hline
D & \begin{tabular}[c]{@{}l@{}}Update sky model (1000)\\ 1000 DI\\ 1000 ionocal\\ 1000 subtracted (DD)\end{tabular} & \begin{tabular}[c]{@{}l@{}}Update sky model (4000)\\ 4000 DI\\ 4000 ionocal\\ 4000 subtracted (DD)\end{tabular} & \begin{tabular}[c]{@{}l@{}}Update sky model (bright sources)\\ 4000 DI\\ Bright sources ionocal\\ 4000 subtracted (DD)\end{tabular} \\ \hline \hline
\end{tabular}
\end{center}
\caption{A summary of different RTS runs investigated. Each cell represents a calibration procedure performed over the whole dataset, the numbers are the amount of sources included in the sky model in the respective calibration step, e.g. `1000 DI' means a direction independent run with a source catalogue comprising 1000 sources. DD here is used to imply the source subtraction step only, separate from the ionospheric correction step denoted by `ionocal'. In row A, only direction independent calibration was done. Row B has both DI and DD, but no ionocal step applied. Row C combines all DI, ionocal and DD, while row D performs an additional calibration iteration with the sky model updated using ionospheric offsets obtained from the first ionocal run. See sections \ref{real_data_results} and \ref{sfinterp} for the discussion motivating the bright sources selection criteria applied in column 3 as well as the results. The analysis strategy in this table was applied to the real data only, but not the simulations}.
\label{tab:rts_real_runs}
\end{table*}

\begin{figure}[!ht]
    \centering
    \includegraphics[width=.4\textwidth]{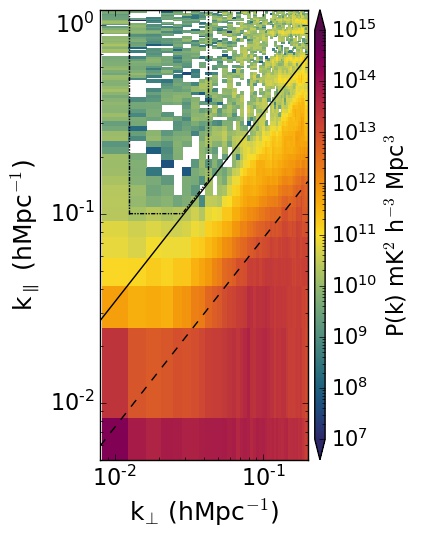}
    \caption{A 2D PS for the best 306 MWA 2-minute observations. The dash-dotted line encloses the modes used for to obtain the 1D spherically averaged PS limits.}
    \label{fig:A1_cleanest}
\end{figure}

We used observations from the MWA Phase $\text{I}$ observed between 2014 and 2015, with the frequency band, frequency and time resolution being identical to the ones of the simulated data. The data is observed in a shift and drift method where the MWA analogue beamformer is electronically pointed to a specific direction (`pointing`) and the sky is allowed to drift overhead for a given observation duration. We use data from the zenith pointing and 2 non-zenith ones, labelled as pointings 2 and 4 \citep[see][]{Beardsley2016}. We select data with a range of ionospheric activity levels, varying from calm to turbulent. However, variations in diffuse emission in the real data would dominate over ionospheric effects, rendering them unquantifiable. We therefore choose observations that were observed within the same LST hour, assuming the variation in the EoR0 field sky temperature over an hour to be negligible ($\sim315$ K at $154$ MHz). In total, we use a set of 452 individual datasets, altogether amounting to $\sim14$ hours of data.

\citealt{jordan2017} introduced a model of categorising ionospheric activities based on the magnitude as well as the isotropicity of the positional offsets obtained during calibration as follows:

\begin{equation}
\begin{split}
    & m < 0.14 \;\; \text{and} \;\; p < 63 	\Rightarrow \text{Type 1} \\
    & m > 0.14 \;\; \text{and} \;\; p < 70 	\Rightarrow \text{Type 2} \\
    & m < 0.14 \;\; \text{and} \;\; p > 63 	\Rightarrow \text{Type 3} \\
    & m > 0.14 \;\; \text{and} \;\; p > 70 	\Rightarrow \text{Type 4},
\end{split}
\label{ion_categories}
\end{equation}

where $m$ is the median ionospheric offsets magnitude across all sources in arcmin and $p$ is the dominant eigenvalue determined by the principal component analysis (PCA) method applied on the ionospheric offset vectors. Thus, $m$ gives a measure of the overall ionospheric turbulence while $p$ serves as a measure of the spatial anisotropy in the ionosphere. For a 2-minute observation processed using the 14 calibration timestamps, the median offset per source was used as the representative offset value for each source. The median offset over all calibrator sources is then used as the model offset value for the observation. In each observation and taking only the brightest $\sim 800$ sources, we used \textsc{Cthulhu} software \citep{jordan2017}, to extract the $m$ and $p$ parameters from the calibration outputs. We then categorised our dataset into  111, 102, 140 and 99 observations for types 1, 2, 3 and 4 respectively. Figure \ref{fig:obsids} summarises the dataset as a function of the ionospheric quality metrics.

Table \ref{tab:rts_real_runs} details the different RTS calibration runs iterated over the dataset. The power spectrum is computed for each RTS configuration integrated over the whole dataset, and where indicated, over each ionosphere type.

\section{Real data results}
\label{real_data_results}
For brevity, runs shown in specific cells in Table \ref{tab:rts_real_runs}, together with their power spectra outputs, are referred to using the row letter and column number, for instance, A1 for the first (1000 sources DI only) cell. Despite the processed data being LST matched, the North-South aligned polarisation is more sensitive to the setting galaxy and therefore shows more power as compared to the East-West dipole, making improvements are less discernible. The main improvement is therefore observed on the East-West polarisation, and this is the polarisation plotted in our results. After examining the calibration solutions, some observations were found to have RFI that had not been excised during the initial flagging process. We chose not to include the affected data in the PS estimation, and only show results for the 306 observations with the best calibration solutions. Figure \ref{fig:A1_cleanest} shows the 2D PS for these 306 observations processed according to procedure A1. The figure shows the expected 2D PS morphology with the foregrounds power concentrated in the wedge while the window has significantly less power. When comparing such 2D spectra from different cells, say A1 and A2, we plot the log of the ratio of the 2 analogous to Figure \ref{fig:k100_ionocal_test} and with a similar interpretation.

The dash-dotted line encloses the $k$-modes within $3.5k_{\parallel}>k_{\perp}$, $0.01<k_{\perp}<0.04$  and $k_{\parallel}>0.1$ which are typically used to obtain the 1D spherically averaged PS limits. This region excludes the entire wedge, as well as avoiding the supra-horizon emission. Additionally, this is a region well sampled by MWA $uv$-coverage. In this work, we apply these exact power cuts as well to compute our real data 1D power spectra.

\begin{figure*}[!th]
    \centering
    \includegraphics[width=\textwidth]{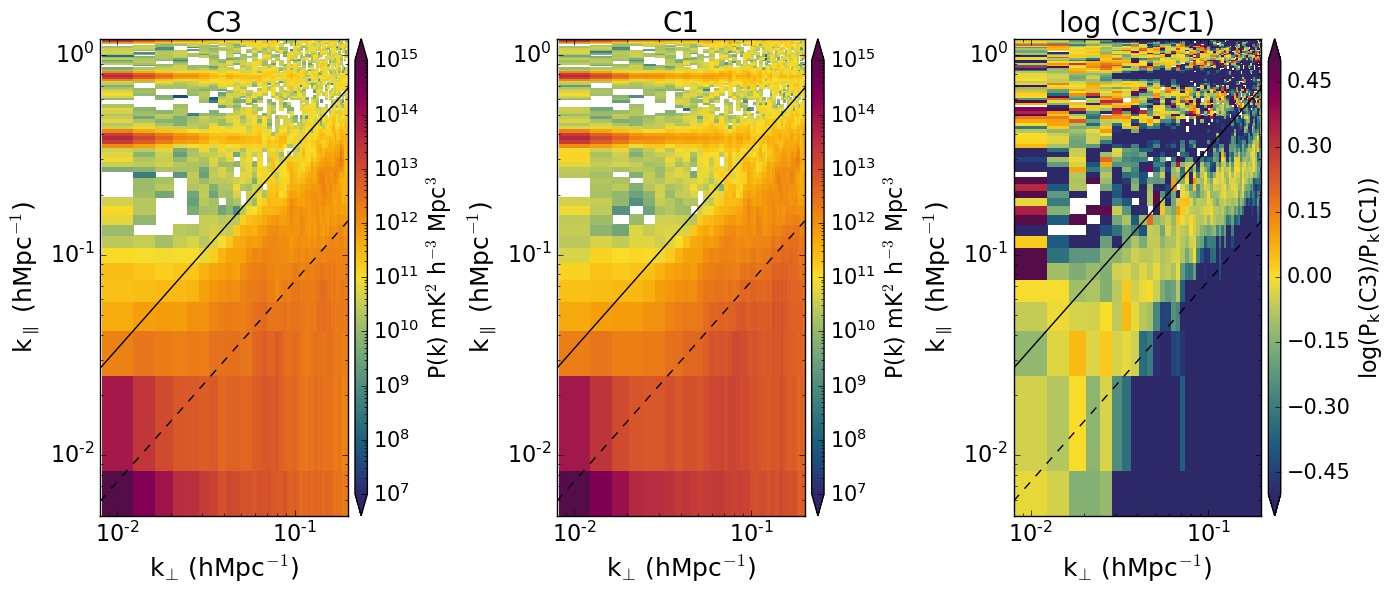}
    \caption{2D PS for the best 37 zenith observations with low ionospheric activity, left: 4000 sky model sources in DI and DD calibration and sources with high SNR corrected for the ionosphere (C3), middle: 1000 sky model sources in DI, ionospheric, and DD calibration (C1), Right: Log of the ratio between first two ($log(C3/C1)$). The bluer colour in the log-ratio plot shows less power in the C3, implying less calibration errors, which in turn reduces contamination in both the foregrounds wedge and the EoR window.}
    \label{fig:C1_vs_C3_yy}
\end{figure*}

\subsection{Overall calibration improvement}
At the time of writing, Cell C1 represents the state of the art for MWA EoR data calibration, as it has the calibration parameters in the current best MWA upper limits \citep{Trott2020, Rahimi2021}. It is already evident from the simulations that this version of calibration is suboptimal
because it is a calibration run with a relatively incomplete sky model and lower foregrounds subtraction. Row C in table \ref{tab:rts_real_runs} investigates whether that can be improved by using run C3. In Figure \ref{fig:C1_vs_C3_yy} which plots $log(C3/C1)$ for low ionospheric activity zenith pointing observations, we show that, in agreement with the simulations, combining a more complete sky model for both DI and source subtraction will result in lower contamination in the EoR window. We can therefore use $\sim4000$ sources in both the DI and the subtraction. Low $k$ modes are expected to have the highest signal-to-noise for the 21 cm signal and therefore the most detectable. As depicted by the bluer region in Figure \ref{fig:C1_vs_C3_yy}, the improvement seen is most enhanced in this valuable low $k$ modes part of the EoR window except for a few grid cells in the lowest $k_\perp$ bin. MWA coarseband harmonics are visible in this figure because the spectra were estimated without CHIPS inpainting of flagged frequency channels. The 2 routines however show similar power along the horizon, as signified by the diagonal feature of  order unity in the ratio plot. This implies the presence of some horizon emission that was not reduced by the improved calibration. Since our calibration sky model in all cases comprised of only sources located within 30 deg of the field centre, bright sources in the beam sidelobes were not removed. This feature is a result of such widefield foregrounds not used in the calibration, as shown in \citep{Pober2016}, as well as potential galactic emission from the horizon. Future works should remove bright sources located up to the horizon in the DD step.

\begin{figure}[ht]
    \centering
    \includegraphics[width=.48\textwidth]{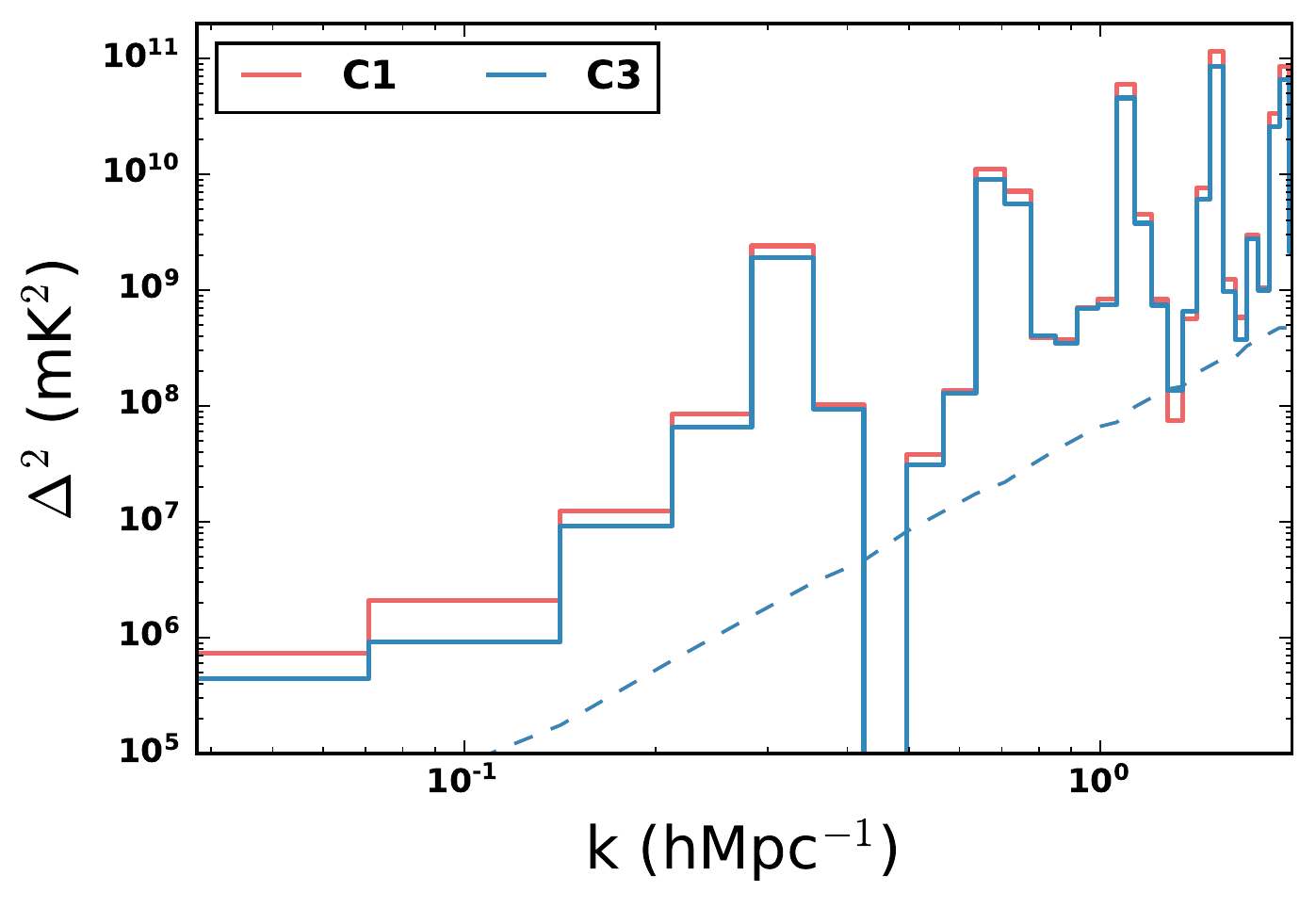}
    \caption{C1 and C3 1D PS for the window modes and zenith pointing data with minimal ionospheric activity. The dashed line corresponds to the thermal noise level.}.
    \label{fig:c1c3cleanestzenith}
\end{figure}

MWA PS results have been found to vary with different sky pointings due to beam modelling errors as well as enhanced ionospheric effects from non-zenith sky directions \citep[e.g.][]{Barry2019}. Figure \ref{fig:c1c3cleanestzenith} shows runs C1 and C3 1D power spectra for the best 37 zenith pointing observations that showed low ionospheric activity (Types 1 and 3 from \citealt{jordan2017}). There is an overall factor of $\sim2$ improvement at the 0.1 hMpc$^{-1}$ scales.
This improvement was consistent at this level for the 3 sky pointings that comprised the analysed observations. We now describe the main factors that contributed to this improvement.

\begin{figure}[!th]
    \centering
    \includegraphics[width=.4\textwidth]{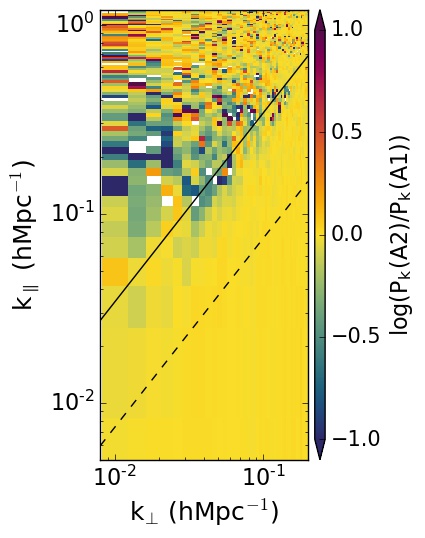}
    \caption{The log of the ratio between the 2D PS for the best observations for a single sky pointing, processed according to cells A2 and A1. In both runs, DD calibration has not been applied. The bluer region in the lower window modes shows improvements from improved DI sky modelling.}
    \label{fig:A2A1_logratio}
\end{figure}

\subsubsection{Sky model completeness}
Row $A$ is used to investigate the usefulness of a more complete sky model for DI calibration by contrasting the PS from a DI run with 1000 (A1) and 4000 (A2) sources in the sky model without any direction dependent calibration or foregrounds subtraction. The increase in number of sources corresponds to a $\sim30\%$ total flux density increase from A1 to A2.
Figure \ref{fig:A2A1_logratio} shows the log of the ratio of cells A2 and A1 2D power spectra integrated over the best observations from one sky pointing. For both runs, the raw data is the same and only DI calibration has been applied. The only difference is that the A2 sky model has a higher completeness level. Modes with a bluer colours represent relative depression of power in the A2 run as compared to A1, while a redder colour would signify relative excess power in A2. Since no foregrounds are subtracted in both cases, the wedge ratio is order unity. However, the window has a clear depression of power, denoted by the bluer colour. This implies improved calibration due to using a sky model that is a more accurate representation of the observed sky.

\begin{figure}[ht]
    \centering
    \includegraphics[width=.4\textwidth]{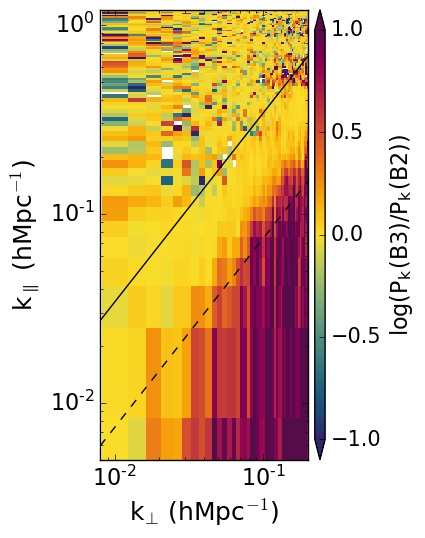}
    \caption{The log of the ratio between the 2D PS for the best 306 MWA 2-minute observations from cells B3 and B2, $log(B3/B2)$. The B3 wedge shows excess power in the small scales as compared to B2. This is because in B3, only $\sim800$ sources with $>1$ Jy flux density were subtracted, and they are less than the 4000 subtracted in B2. The window ratio is noise-like.}
    \label{fig:B3B2_logratio}
\end{figure}

\begin{figure}[!ht]
    \centering
    \includegraphics[width=.4\textwidth]{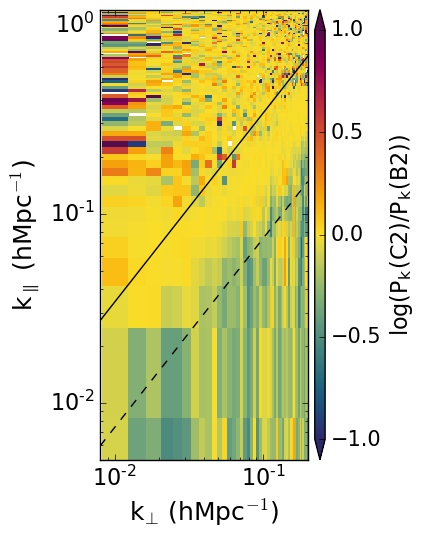}
    \caption{The log of the ratio between the 2D PS for the best 306 MWA 2-minute observations from cells C2 and B2, $log(C2/B2)$. The only difference in the two runs is that C2 has ionospheric correction while B2 has not. The C2 wedge shows less power as compared to B2, but despite this, the EoR window remains noise-like. We can conclude that the ionospheric correction is still indirectly advantageous, as it reduces the power level that can leak into the EoR window  as a result of any other systematic, as shown in Figure \ref{fig:k100_ionocal_test}.}
    \label{fig:C2B2_logratio}
\end{figure}

\subsubsection{Foregrounds subtraction level}
Row B was run to contrast the effect of the number of sources subtracted without any ionospheric calibration applied. As expected, the B2 and B3 window were found to have less contamination than B1, primarily due to the more complete DI sky model used. However, in Figure \ref{fig:B3B2_logratio}, the B3 wedge shows excess power as compared to B2. This is because in B3, only $\sim800$ sources with $>1 Jy$ flux density were subtracted, and they are less than the 4000 subtracted in B2. The window, however, still looks noise-like. Since the simulations suggest that we expect to see less window contamination as well, we attribute this noise-like behaviour to lack of enough sensitivity as well as the fact that no ionospheric correction has been applied yet. To confirm this, we present Figure \ref{fig:C2B2_logratio}, $log(C2/B2)$, which tests for the effect of ionospheric correction. The wedge from C2 shows lower power. Despite, the EoR window still remaining noise-like, we can conclude that the ionospheric correction is still indirectly advantageous as it reduces the power level that can leak into the EoR window as a result of any other systematic.

As found in the simulation results, the performance of RTS ionospheric correction deteriorates at extreme ionospheric activity. This can lead to inaccuracies during source subtraction and, potentially, signal loss. Such over-subtraction was found in residual images for observations with active ionosphere (types 2 and 4). No significant over-subtraction was observed in low ionospheric activity (types 1 and 3, Figure \ref{fig:c1c3images}) images, except for sources with incorrect sky models.

\subsubsection{Application and accuracy of ionospheric correction}
\label{sigmafluxcut}
\begin{figure}[ht]
    \includegraphics[width=.5\textwidth]{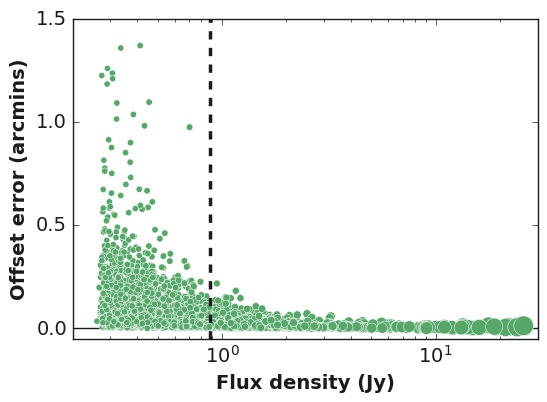}
    \caption{Position offset errors from the RTS ionospheric offsets estimation as a function of source brightness. The larger dots represent sources with higher SNR. The position errors increase with lower SNR. The dashed line represents a qualitative flux threshold chosen at the `elbow' of the trend, for categorising 'faint' and 'bright' sources during calibration.}
    \label{fig:faint_source_errors}
\end{figure}

\begin{figure}[ht]
    \centering
    \includegraphics[width=.4\textwidth]{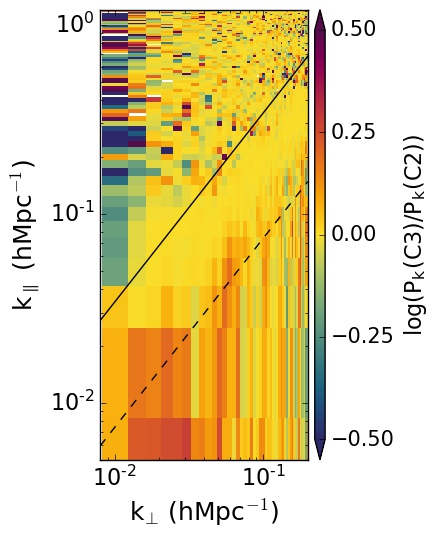}
    \caption{$\text{log}(C3/C2)$. The noise-like window implies that we do not see significant differences in the window based on the number of sources that ionospheric correction is applied to. The ionospheric correction to fewer sources in C3 is however apparent, signified by the apparent redness in the wedge.}
    \label{fig:C3C2_logratio}
\end{figure}

By increasing the number of calibration sources, we are including fainter sources in our sky model. These sources have an inherently larger positional inaccuracy, which is propagated to the RTS-derived ionospheric offsets. Their inclusion in ionospheric calibration could be more detrimental than not using them at all.
How faint we can allow a source to be in order for it to be included in the ionospheric calibration is a pertinent question. We investigate this effect by comparing pairs of simulated 2-minute observations, identical in all aspects including the ionosphere except that one has no thermal noise included. The ionospheric S-screens were used with magnitude hyperparameters of 10, 25, 50 and 100 (maximum median offset of 0.2 arcmin).
Figure \ref{fig:faint_source_errors} shows how position errors evolve as sources get fainter for the s50 simulation, with the accurate offsets being the RTS offsets from the simulation with lower noise. Noise confusion results in higher RTS position errors for the faint sources. The RMS level from the central 10deg by 10deg region of a 2-minute snapshot image made from the visibilities with noise using robust 0 Briggs weighting was found to be $\sim 5$ mJy$/$beam. This noise level is consistent with the predicted MWA snapshot thermal noise (e.g., \citealt{Wayth2015}). The observed trend in Figure \ref{fig:faint_source_errors} was consistent in the four turbulence levels used, and we found a maximum percentage error of between $\sim5-20\%$.

These simulations are of course optimistic, real data might be dominated by different sidelobes and other confusion sources.
We use a conservative $\sim1$ Jy flux threshold to categorise between bright and faint sources in the ionospheric correction of the real data runs in the final column of table \ref{tab:rts_real_runs}. This cutoff is qualitative, chosen at the `elbow' of the trend observed in the simulated offset errors. We note however that the RTS does average over multiple channels during ionospheric correction, and the $\sim1$ Jy cutoff would result in bright sources with sufficient signal-to-noise ratio regardless of the dominant noise cause.

We obtained the system temperature as the sum of the recorded sky temperature for the respective observation with $\sim50 \text{K}$ receiver temperature for the MWA \citep{Tingay2013, Wayth2015}. 
Similar to the simulation, the chosen $\sim1$ Jy flux threshold corresponded to a $\sim 2 \sigma$ SNR for a single channel at the centre of the frequency band (154 MHz) over 8s duration. This threshold resulted in $\sim800$ sources that we deemed bright enough to be corrected for the ionosphere without introducing errors. 

The faint sources can however still be subtracted out based on their catalogue positions. This is the procedure applied in cell C3 (the best calibration run). C2 is expected to have ionospheric correction errors from sources with a low signal-to-noise ratio (SNR). From Figure \ref{fig:C3C2_logratio}, C3 is seen not to have a significant difference from C2 in the window.

\subsection{Results per ionosphere type}
From the different categories of ionospheric activity described, the real data did not show conclusive differences in the power spectrum. We attribute this to the presence of other dominant systematics \citealp[e.g.,][]{Trott2020}. The major sources of these systematics are errors in the models of the instrument and the sky (the diffuse component of the sky not being included in the calibration model), as well as errors inherent to the calibration and power spectrum estimation algorithms. The effect of active ionosphere is, however, already clear from the simulations.
\citealt{Trott2018} uses analytical models to show how ionospheric spatial structures would introduce biases in the power spectrum, for various reasons that make any direct comparison of real data runs from different ionospheric categories unfair, a result similar to theirs is not observed in this work.

\section{Advanced ionospheric modelling and correction methods}
\label{sfinterp}
\subsection{Iterative sky model corrections based on ionospheric offsets}
\label{s_feedback}
This section addresses row D in table \ref{tab:rts_real_runs}.
We can potentially improve the calibration of an observation by performing two calibration iterations. After the first iteration, we obtain the offset of each source from its catalogued position and make a modified catalogue that includes the shift. This modified catalogue provides a more accurate description of the measured visibilities for that observation and should result in more accurate calibration gain solutions. \citealt{Yoshiura2021} performed this procedure for MWA ultra-low data, where the ionospheric impact is much higher than at the MWA EoR low band. Here, we investigate whether this method is viable in the low band.

\begin{figure}[!t]
    \centering
    \includegraphics[width=.5\textwidth]{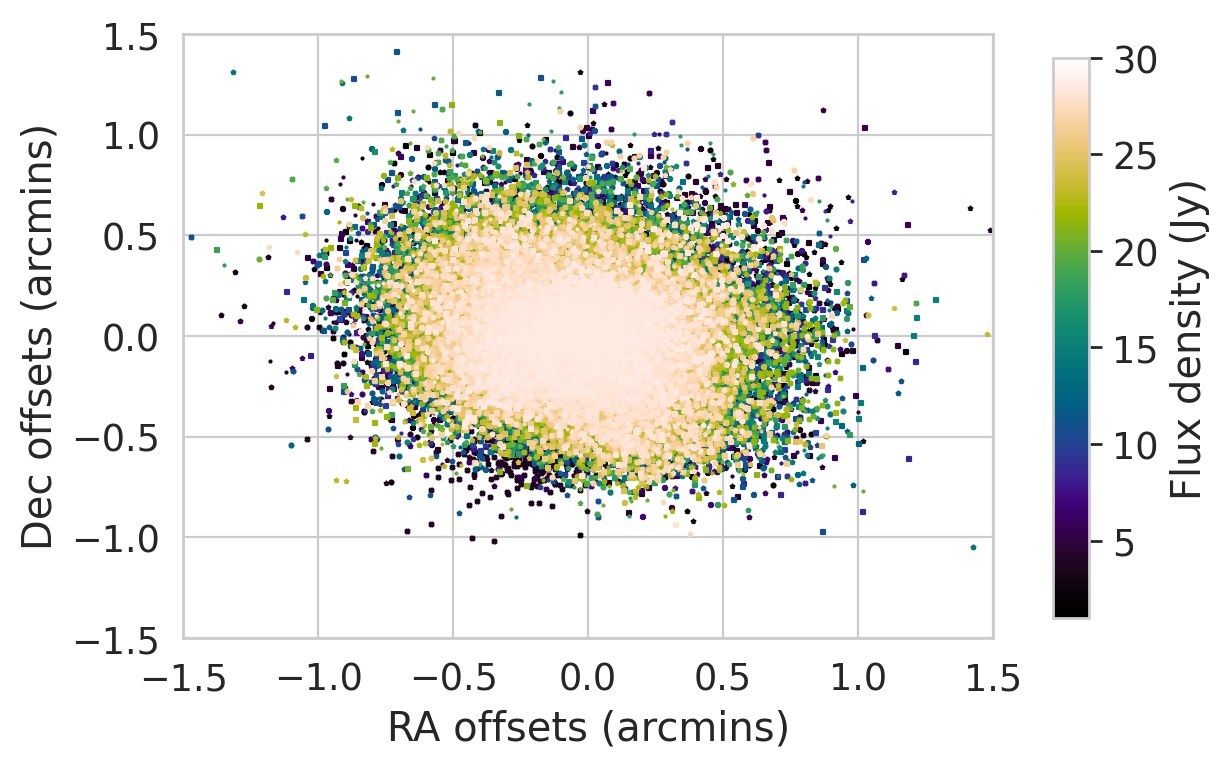}
    \caption{Offsets distribution per source over all observations. Such a Gaussian-like distribution is expected for purely ionospheric offsets.}
    \label{fig:offsets_over_all_obsids}
\end{figure}

In order to perform such ionospheric correction procedures accurately, we need to ascertain that the observed source positional offsets are predominantly ionospheric and not caused by other systematics. We test this by obtaining a list of common sources in all the 452 datasets and investigating the distribution of the offsets from each individual source. The offset magnitude per source is assumed to be static over the 2 minutes dataset duration, and is approximated by the median of the offset magnitudes obtained for the 14 calibration timestamps. Ionospherically induced offsets on a source caused by multiple random ionospheric conditions at different times should be normally distributed around zero. Any other distribution centred at a different value implies a systematic error in the catalogued position of the sources. Figure \ref{fig:offsets_over_all_obsids} shows the distribution of offsets obtained for the total 452 observations. Except for the outliers, there is no distinctive disparity of the source offsets from the expected distribution, signifying lack of significant systematics.

Similarly, the ionospheric calibration procedure is not expected to affect the amplitude gain solutions obtained during prior calibration. We obtained the scaling factors applied to the amplitude gains for each source over all the datasets. A scaling factor distant from unity would signify a systematic in the properties of the calibrator source. 
The amplitude scales showed a maximum spread of $\sim 20\%$ around unity, implying no major systematics in the source flux densities are introduced by the ionospheric calibration.

\begin{figure}[!t]
    \centering
    \includegraphics[width=.4\textwidth]{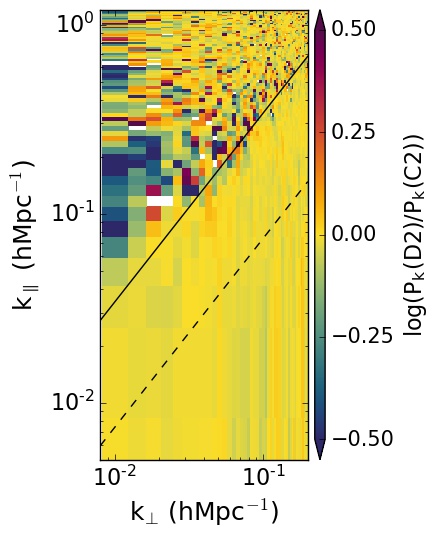}
    \caption{The log of the ratio between the 2D PS for cells D2 and C2, $log(D2/C2)$ obtained for pointing 2 data. There is marginal improvement observed in the wedge and the window.}
    \label{fig:D3C3_logratio}
\end{figure}

After this correction, there were minimal improvements observed across all modes, but varying across the 3 sky pointings. The most improvement was seen in pointing 2 with the possible reason for this being that pointing 2 had more datasets as compared to the other 2 pointings and also has less galactic contamination \citep{Beardsley2016}. Figure \ref{fig:D3C3_logratio} shows the log of the ratio between the 2D PS from cells D2 and C2; $log(D2/C2)$ from the pointing 2 data. The figure shows that the sky model correction applied does indeed reduce ionospheric contamination, albeit at a minimal level, which at the current EoR calibration precision levels, can easily get dominated by other systematics.

\begin{figure*}[!th]
    \centering
    \includegraphics[width=\textwidth]{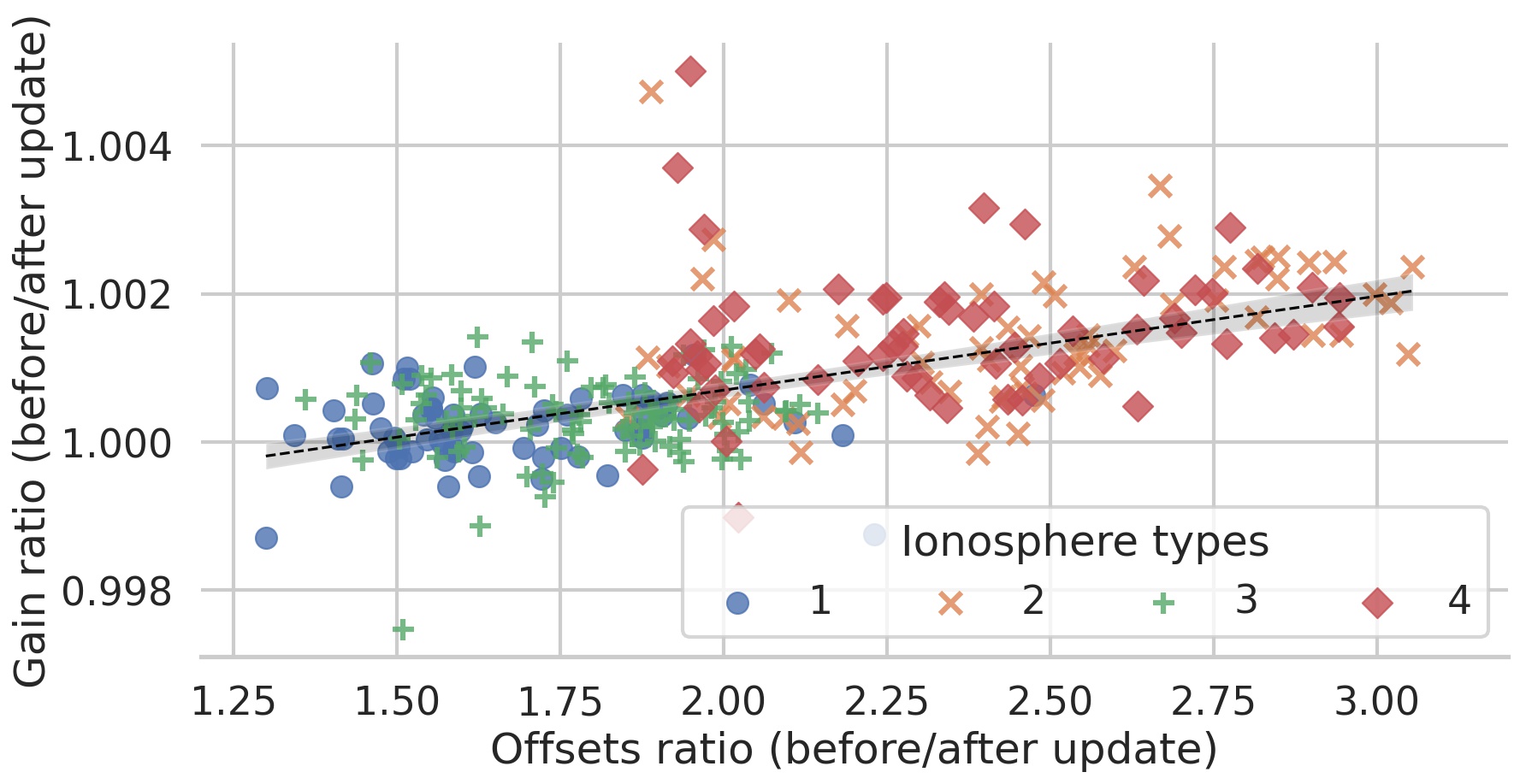}
    \caption{Ratio of the source position offsets and the gains amplitude before and after the sky model updates. The markers represent the 4 ionospheric categories, while the black dashed line shows the correlation trend with a $99\%$ confidence interval (grey shaded region). The trend shows a positive correlation between ionospheric activity and both gains amplitudes and position offsets. Updating the sky model has a much higher impact on the offsets (up to a factor of 3 reduction) as compared to the visibility amplitudes ($<1\%$ reduction).}
    \label{fig:offsets_amps_ratio_plot}
\end{figure*}

To further investigate the minimal improvement observed, we present Figure \ref{fig:offsets_amps_ratio_plot} which shows the ratio of the source position offsets and the gain amplitudes before and after applying sky model updates. The figure shows a clear positive correlation between ionospheric activity with both the gains amplitude and position offsets.
The updated sky model results in a factor of $\sim1.2$ to $3$ (up to $\sim67\%$) reduction in position errors. However, the maximum change in the amplitudes is less than $1\%$. This variation in the magnitude of the two effects is direct evidence to the well known dominance of ionospheric first order effects (refraction phase errors) over visibilities amplitude scintillation; a higher order ionospheric effect.

A challenge in the current application of this correction is not taking the spectral nature of ionospheric offsets into account when correcting the sky models.
A full treatment would require an offset-corrected sky model for every individual frequency channel, a requirement that is not feasible with the current tools.

\subsection{Ionospheric modelling with interpolation methods}
\begin{figure*}[t!]
    \centering
    \includegraphics[width=\textwidth]{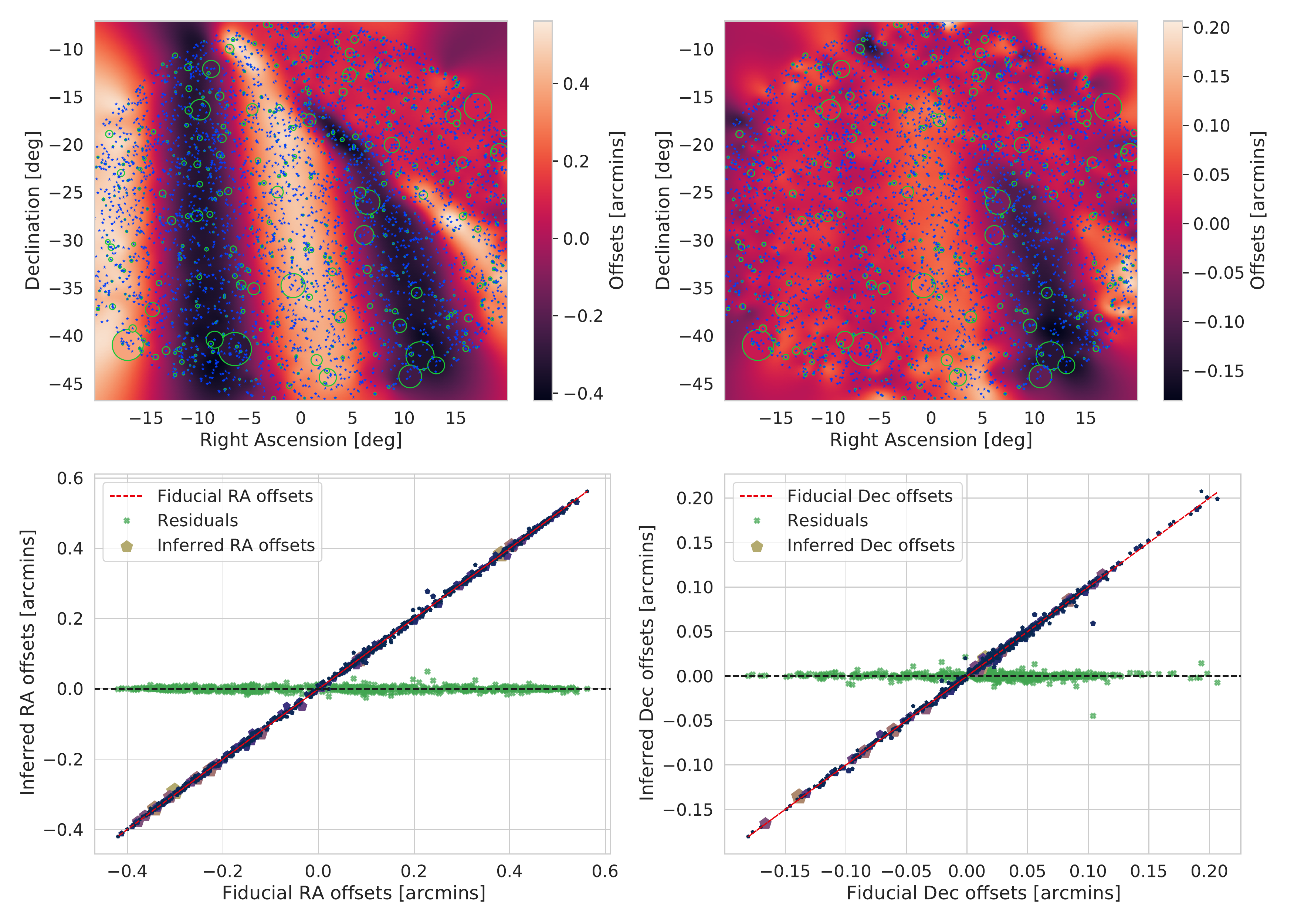}
    \caption{Top: Ionospheric spatial structure interpolated using the differential RA (left) and Dec (right) offsets for an EoR0 \textsc{sivio} simulation over the main lobe of the MWA. Bottom: The recovered RA and Dec offset values for the interpolants with their residuals. The fiducial offsets are the ones measured during DD calibration. Interpolation corrects spurious ionospheric gains obtained for low SNR sources during DD calibration.}
    \label{fig:interpolated}
\end{figure*}

In order to minimise ionospheric calibration errors associated with the faint sources, we attempt to interpolate the ionospheric offsets over the field of view using the offsets from high SNR sources as the interpolants. We use the radial basis functions (RBF) 2D interpolation method provided by the \textsc{SciPy} library. We note that this method is not unique and other methods such as Karhunen-Lo{\`{e}}ve base functions as well as Gaussian processes regression and have been applied in different ionospheric calibration use cases e.g. \citep{Intema2009, Hurley-Walker2018, Albert2020a, Albert2020b}. We expect this method to recover accurate offsets for active ionopheric conditions up to the limit where RTS calibration fails due to extreme position offsets. Figure \ref{fig:interpolated} shows the interpolated RA (left) and Dec (right) offsets from simulated visibility data with s50 ionospheric parameters. The green circles are centred at the brightest sources in the field, whose offsets were used for the RA and Dec offsets interpolation. The blue dots are all the remaining fainter sources in the sky model. The bottom panel shows inferred values of the interpolants (dark polygons) compared with their fiducial offset values (solid red line) and the residuals are shown in green. The fiducial simulated ionospheric structure is well recovered, resulting in minimal residuals, and we conclude that this sufficiently corrects for the spurious noise-confused offsets obtained for the fainter sources by the calibration algorithm.
However, no significant improvement was observed on the PS as a result of this smoothing procedure when compared to the D2 and D3 runs.

\section{Discussion}
\label{sec:discuss}
We have shown that we can obtain lower power contamination in the MWA lowband EoR window. This has been possible due to three main factors:
\begin{enumerate}
\item Improved computational resources.
\item Higher sky models completeness levels.
\item Rigorous ionospheric correction and foreground subtraction.
\end{enumerate}

The DI step alone provides the biggest improvement. This is in line with other literature in the field, which have shown that all kinds of calibration (e.g., sky-based and redundant calibration) suffer from sky model incompleteness errors \citep{Byrne2019, Barry2016, Patil2016}. Foreground subtraction of more sky sources combined with increased ionospheric correction in the DD step accounts for more contamination reduction.

To avoid errors associated with sources confused by the thermal noise level, we apply the optimum flux density threshold for the RTS ionospheric correction based on the recorded per observation parameters. However, the errors introduced by applying ionospheric corrections to low SNR sources are found not to be significant. These errors might not observable as a result of the limited amount of data integrated, and future more sensitive integrations could unearth the errors. Nevertheless, this would be an especially important factor to consider for future deeper analyses.

Using ionospheric calibration information for improved sky modelling has shown to marginally improve the PS.
Similar results are observed when  we apply the interpolation method as a solution for mitigating the faint source errors. Alternative interpolation methods for ionospheric modelling can also be examined similarly, and this has been done for different use-cases with Gaussian Process Regression (GPR) methods and others \citep{Intema2009, Albert2020a, Albert2020b}. The improvements observed are, however, sky pointing and dataset dependent, implying that the improvement achieved is at a level dominated by other systematics such as instrumental beam modelling errors. This step involves an additional calibration iteration, but it would still be worthwhile to apply the procedure to larger datasets in future when EoR limits are much closer to the 21 cm signal level. Furthermore, \cite{Trott2018} showed the bias imprinted by the ionosphere to the power spectrum of the 21 cm signal. They find that correction of the ionosphere effects both before source subtraction and afterwards in the residuals is key to getting rid of this bias.

Our additional ionospheric correction process does not account for the variation of the offsets over the frequency band pass. This would require use of different sky models per frequency channel, which is not supported by the RTS. Actual development of the RTS capabilities is beyond the scope of this work. However, future calibration algorithms that aim to fully correct for the ionosphere while minimising spectral errors should take this process into account.  Despite this, based on the marginal improvements observed by applying the additional ionospheric corrections,  we conclude that a stringent calibration such as the one done in the C3 run is sufficient, and the ionosphere is not a showstopper for EoR science at frequencies above 100MHz. This is in agreement with previous literature \citep[e.g.][]{Vedantham2016}.

The approach of calibration using select sources from composite catalogues of the target EoR field of view is not unique to the MWA. Neither is the need for sufficient SNR from the number of sources used as well as the correction
of direction dependent errors while attempting to minimise computation costs, see e.g. \citealt{Mertens2020, Patil2017, Yatawatta2016} for similar endeavours as applied to the LoFAR EoR experiment. This work adds to such efforts targeting to achieve improved calibration for MWA EoR analysis with the currently available resources.

Any EoR calibration routine needs to ensure that the fidelity of the underlying 21 cm signal is not compromised. Lack of a diffuse emission component in the DD calibration step can be a cause of signal loss \citep{Patil2016}. However, in our analysis, the expected loss should be insignificant since,  as earlier mentioned, we do not perform peeling in its original sense. The point source ionospheric treatment also should affect different scales to those of the 21 cm signal. Nevertheless, the use of the recommended calibration strategy from this work for EoR limits in future will incorporate an end-to-end signal loss analysis.

Obtaining the best performing calibration routine using the available RTS/CHIPS MWA pipeline was one of the main aims of this paper. After finding that the iterative calibration for ionospheric correction provides only marginal improvement to the PS in the frequency range of this work, we recommend the C3 and C2 calibration runs as the current most optimum strategies.
The enhancements in those runs result in a $\sim2$ factor improvement in the EoR window PS. Future work will involve applying this strategy to a larger MWA dataset for improved PS limits. This work not only provides an improved EoR calibration strategy, but also contributes to the need for end to end pipeline verification, which is getting stronger as the EoR science community gets closer to detecting the signal.

\begin{acknowledgements}
Kariuki Chege thanks Siyanda Matika for valuable discussions during this work. This work was supported by the Centre for All Sky Astrophysics in 3 Dimensions (ASTRO3D), an Australian Research Council Centre of Excellence, funded by grant CE170100013. CMT is supported by an ARC Future Fellowship through project number FT180100321. SY is supported by JSPS KAKENHI Grant No 21J00416 and JSPS Research Fellowships for Young Scientists. The International Centre for Radio Astronomy Research (ICRAR) is a Joint Venture of Curtin University and The University of Western Australia, funded by the Western Australian State government. This scientific work makes use of the Murchison Radio-astronomy Observatory, operated by CSIRO. We acknowledge the Wajarri Yamatji people as the traditional owners of the Observatory site. Support for the operation of the MWA is provided by the Australian Government (NCRIS), under a contract to Curtin University administered by Astronomy Australia Limited. We acknowledge the Pawsey Supercomputing Centre, which is supported by the Western Australian and Australian Government.
\end{acknowledgements}

\bibliographystyle{pasa-mnras}
\bibliography{bibli}

\begin{appendices}
\renewcommand\thefigure{\thesection.\arabic{figure}}    
\setcounter{figure}{0}

\section{Residual images from C1 and C3 procedures}
\begin{figure*}
     \centering
     \begin{subfigure}[b]{1\textwidth}
         \centering
         \includegraphics[width=.8\textwidth]{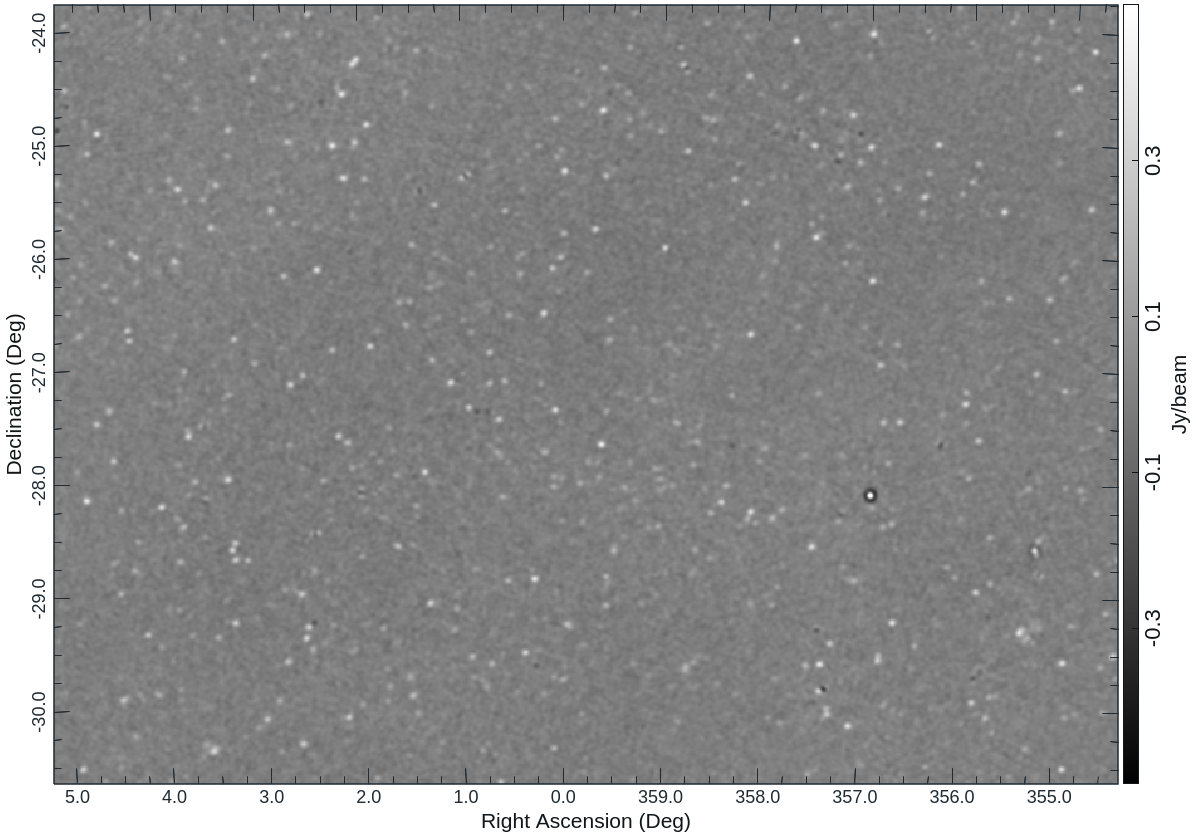}
         \caption{C1}
         \label{c1image}
     \end{subfigure}
     \begin{subfigure}[b]{1\textwidth}
         \centering
         \includegraphics[width=.8\textwidth]{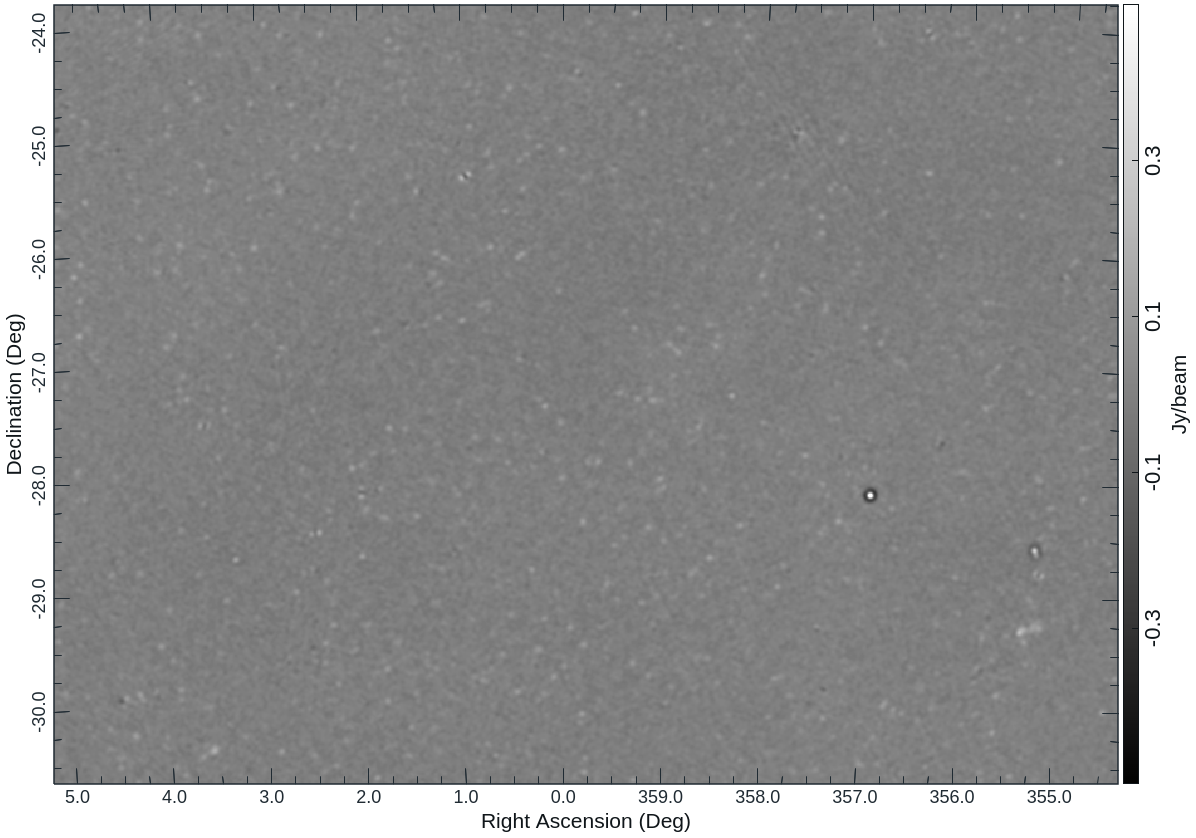}
         \caption{C3}
         \label{c3image}
     \end{subfigure}
        \caption{C1 (top) and C3 (bottom) residual images after subtraction of 1000 and 4000 sources respectively.}
        \label{fig:c1c3images}
\end{figure*}

Over-subtraction of sources due to inaccurate sky models or inaccurate ionospheric source offset corrections can lead to power loss. Figure \ref{fig:c1c3images} shows the central region of residual images made from visibilities calibrated using procedures C1 (\ref{c1image}) and C3 (\ref{c3image}). Both images are made from the same 2-minutes observation carried out on 14th November 2014 and using the same imaging parameters. No over-subtraction is observed in both images. 

As expected, the C3 image shows lower source residual amplitudes than the C1 image.
A bright poorly subtracted source is also conspicuous on the right lower quadrant of both images, and the new improved source catalogue by \citep{Lynch2021} will enable more accurate subtraction of sources.

\section{S-screen analogue of Figure 5}

\begin{figure*}[!ht]
    \centering
    \includegraphics[width=\textwidth]{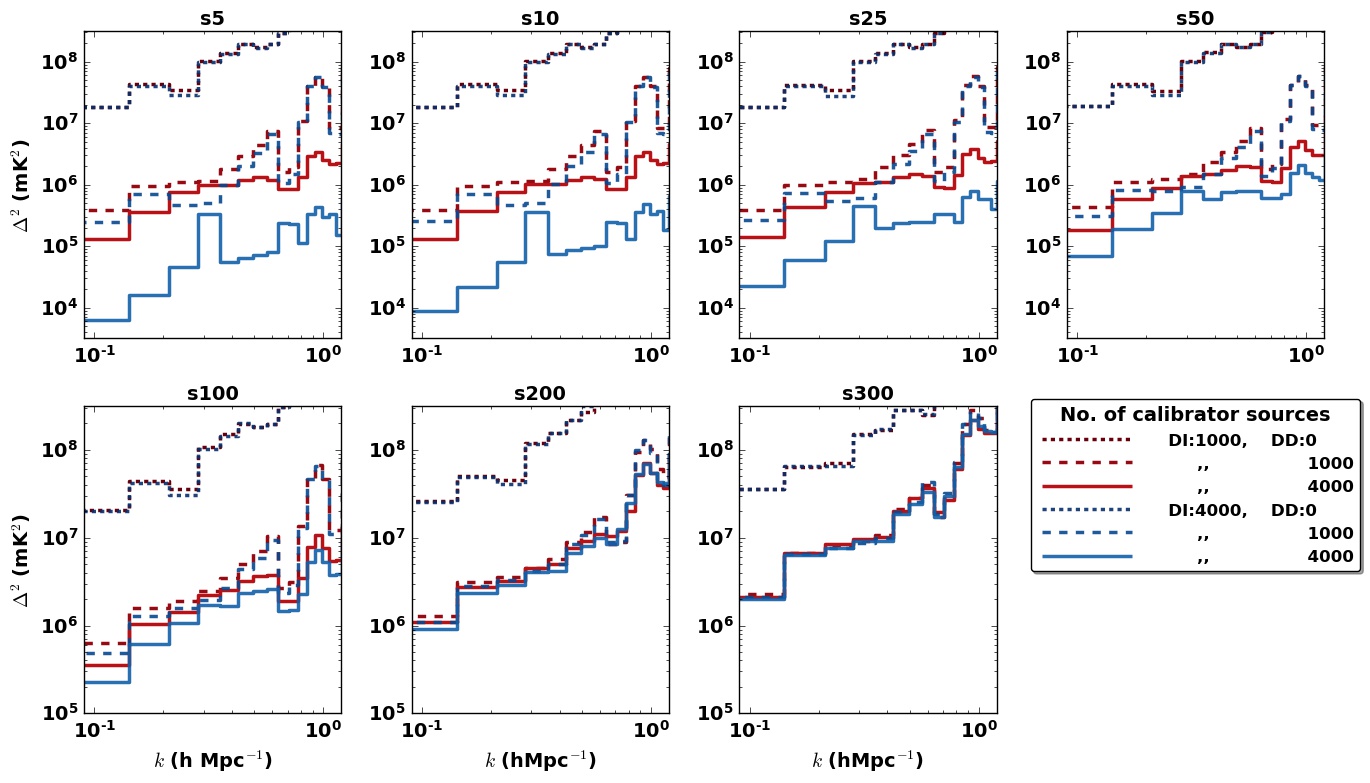}
    \caption{Similar to Figure \ref{fig:ptvspl_k} but using ionospheric S-screen}
    \label{fig:ptvspl_s}
\end{figure*}

\end{appendices}

\end{document}